\documentclass{interact}

\usepackage{upgreek}
\usepackage{caption}
\usepackage[font=footnotesize,labelfont=bf]{caption}
\usepackage{tabularx}
\usepackage{dcolumn}
\usepackage{booktabs}
\usepackage{placeins}
\usepackage[numbers,sort&compress]{natbib}

\usepackage{epsfig}
\usepackage{url}
\usepackage{graphicx}
\usepackage{subcaption}
\usepackage{amsmath}
\usepackage{indentfirst}
\usepackage{mathrsfs}
\usepackage{booktabs}
\usepackage{setspace}
\usepackage{enumerate}
\usepackage{dsfont}

\usepackage{multirow}
\usepackage{adjustbox}
\usepackage{footmisc}

\usepackage{comment}
\usepackage{bm}
\usepackage{mathtools}
\usepackage{bbm}
\usepackage{xcolor}
\definecolor{NavyBlue}{RGB}{43,140,190}
\usepackage[colorlinks=true,linkcolor=NavyBlue,citecolor=NavyBlue,urlcolor=NavyBlue]{hyperref}

\begin{document}
\title{Multiple Outlier Detection in Samples with Exponential \& Pareto Tails }
\author{
\name{Didier Sornette\textsuperscript{a,b}\thanks{Emails: dsornette@ethz.ch(corresponding author); ranwei@ethz.ch} and Ran Wei\textsuperscript{c}}
\affil{\textsuperscript{a}Southern University of Science and Technology, Institute of Risk Analysis, Prediction and Management, Shenzhen, China; \\
\textsuperscript{b}Swiss Finance Institute, c/o University of Geneva, 1211 Geneva 4, Switzerland; \\
\textsuperscript{b}ETH Zurich, Department of Management, Technology and Economics, Zurich, Switzerland}
}

\maketitle

\begin{abstract}
We introduce two ratio-based robust test statistics, \emph{max-robust-sum} (MRS) and \emph{sum-robust-sum} (SRS), designed to enhance the robustness of outlier detection in samples with exponential or Pareto tails. We also reintroduce the inward sequential testing method -- formerly relegated since the introduction of outward testing -- and show that MRS and SRS tests reduce susceptibility of the inward approach to masking, making the inward test as powerful as, and potentially less error-prone than, outward tests. Moreover, inward testing does not require the complicated type I error control of outward tests. A comprehensive comparison of the test statistics is done, considering performance of the proposed tests in both block and sequential tests, and contrasting their performance with classical test statistics across various data scenarios. In five case studies -- financial crashes, nuclear power generation accidents, stock market returns, epidemic fatalities, and city sizes -- significant outliers are detected and related to the concept of `Dragon King' events, defined as meaningful outliers that arise from a unique generating mechanism. 
\end{abstract}



\section{Introduction}\label{sec:Intro}

Anomalous observations, while sometimes dismissed as nuisances, can be of primary relevance in applications ranging from medical diagnosis to climate science (see \cite{Aggarwal} for examples). The statistical analysis of these outliers has been extensively documented in the literature (e.g., \citep{Barnett1994,Hawkins1980} are classic references). Traditionally, statistical methods focuses on testing outliers relative to a \emph{null} model, i.e., a model without outliers, which in many cases assumes an underlying Gaussian distribution.
 Common approaches include the contaminated normal models that assume Gaussian sample with Gaussian outliers (Sec. 3.4 of \cite{Hawkins1980}), and Gaussian mixture models that are often used to detect anomalous sub-populations \citep{hodge,aitkin}. However, empirical data in many fields often do not follow Gaussian distributions. In many scenarios, data can be better described by distributions with fatter tails such as exponential or power law distributions. In particular, Pareto (power law) distributions are prevalent across a broad spectrum of phenomena, ranging from natural hazards like earthquakes, landslides, floods, and tsunamis, to industrial catastrophes such as chemical spills, nuclear accidents, and power blackouts, and extend to social systems and geopolitical events, including the distribution of wars and conflicts intensities measured by human losses \citep{LaherSornette,Mitzenmacher04,NewmanMEJ05,Sornette2006}.
 
This paper thus focuses on the detection of outliers in samples having approximately exponential or Pareto tails. Notably, through a simple transformation, outlier tests designed for exponential samples can also be applied to Pareto samples. This places the test statistics with exponential underlying distribution at the center of our study. Moreover, Extreme Value Theory (EVT) suggests further generality of the exponential distribution by providing that general `well-behaved' distribution functions asymptotically exhibit either exponential or Pareto tails \citep{Embrechts}. Although the exponential null models have been covered in the literature (\citep{Bala} provides a review), we show that its application extends far beyond common usage. In addition, the tests we consider are independent of the parameter of the exponential distribution function, avoiding the danger of estimating the parameters of the distribution function in the presence of outliers, which can lead to strong biases.

Grounded in classical testing approaches, we introduce the two ratio-based test statistics, \emph{max-robust-sum} (MRS) and \emph{sum-robust-sum} (SRS), which are respectively modifications of the \emph{max-sum} (MS) \citep{Kimber1982} and the \emph{sum-sum} (SS) test statistics \citep{Lewis1979, Chikka1983}. The modifications involve altering the total sum in the ratio to a partial sum, ensure `robust' detection of outliers by recalibrating the weight given to potential outliers in the calculation.
 
Historically, block testing and sequential testing have been utilized in outlier detection. The masking and swamping issues in block testing have led to the development of sequential testing. While inward sequential tests are particularly vulnerable to masking, outward sequential tests were developed to mitigate this issue \citep{Rosner1975,Kimber1982}, and have since become the standard approach. Despite their widespread adoption, outward tests are substantially more complicated as they require multiple testing corrections that control the Type I error rate. In this study, we show that the MRS and SRS test statistics effectively addresses the masking problem of inward tests, making them competitive with outward tests. Our method opens new avenues for using inward tests without the extensive complications associated with outward methods. Furthermore, we conduct a comprehensive comparison of different statistics performed under these testing approaches, providing useful practical insights that -- in the opinion of the authors -- go beyond the existing literature (e.g., \cite{LinBala2009,Chikka1983,Balasooriya1994,LinBala2014}).

We also offer five interesting and extensive case studies including financial crashes, nuclear power generation accidents, stock market returns, epidemics fatalities, and city sizes. These studies shifts the outlier detection from reliability/failure applications (the exponential case) towards applications in risk modeling (the Pareto case). A number of studies have found suggestive evidence that there are extreme events `beyond' the Pareto sample \citep{SornetteDK2009,SorOuiDrag}. This brings into play the concept of `Dragon Kings' \citep{SornetteDK2009}. We show that the proposed outlier detection method effectively identifies these `Dragon Kings'.
  
The paper is organized as follows. Section 2 proposes the test statistics and provides their analytical distribution functions. Section 3 presents a variety of comparative studies across different scenarios, evaluating the performance of various tests with both dispersed and clustered outliers and their susceptibility to masking and swamping. Section 4 describes the general methodology in outlier detection and the argument, based on EVT, that supports the generality of the exponential outlier test. Section 5 explains the Dragon King (DK) concept and gives a case study on financial market crashes. The Appendix presents four additional case studies that highlight the findings from previous sections. Section 6 concludes.

\section{Test statistics }
  
The setup is an ordered sample $x_{(1)}>x_{(2)}>...>x_{(n)}$ where $n-k$ of the observations are i.i.d. realizations of a random variable, $X\overset{\text{iid}}{\sim}\text{Exp}(\alpha)$, with the distribution function,
  \begin{equation}
   F_X(x)=1-\text{exp}\{-\alpha x \},~x\geq0,~\alpha>0~,
   \label{eq:Exp}
  \end{equation}
and the remaining $k$ points are outliers, also i.i.d. with some different distribution function, and independent of $X$. It is unknown which points are outliers, and the objective is to detect them. Moreover, if $X$ is exponentially distributed, then $Y=u \exp\{X\}\overset{\text{iid}}{\sim}\text{Pareto}(\alpha,u)$. That is, the exponential of an exponential random variable has the Pareto distribution function,
  \begin{equation}
   F(x)=1-(x/u)^{-\alpha},~x \geq u,~\alpha>0~.
   \label{eq:Pareto}
  \end{equation}
  
 Therefore, one can take the logarithm of Pareto samples and apply outlier tests intended for exponential samples.

 \subsection{Gallery of test statistics}\label{sec:gallery}

We now propose the MRS and SRS statistics, and review other standard test statistics for outlier detection in exponential samples. In general, outlier test statistics compare the `outlyingness' of the suspected outliers by contrasting them against some measure of dispersion within another subset of the data. Some of the measures are based on spacings or maxima, others on the sums of observation sizes.
  
 The \emph{max-robust-sum} (MRS) statistic for the $j$-th rank,
   \begin{equation} 
   T_{j,m}^{MRS}= \frac{ x_{(j)} }{ \sum_{i=m+1}^{n} x_{(i)} },~~m\geq 0~,
   \label{eq:MRS}
   \end{equation}
 is a modification of a classic statistic \citep{Kimber1982}, here referred to as \emph{max-sum} (MS) statistic. $m$ is a pre-specified maximal number of outliers, and the MS statistic is recovered when $m=0$. Index $j$ allows the test to be used in sequential procedures for $j=1,...,m$. Having $m>0$ in the denominator prevents masking: when the true number of outliers is $r>0$, for $m<r$, there will be $m-r$ outliers in the denominator that will make $x_{(j)}$ appear less outlying. This consideration becomes crucial in inward testing, and is similar to using robust scale estimates in the case of outliers relative to a normal population \citep{Iglewicz1982}.  Thus, the choice of $m$ is a tradeoff between sample size (power) and sample purity (masking avoidance). The classic MS statistic has optimal properties in the presence of a single outlier \citep{Hawkins1980} as it uses a single value in the numerator. MRS/MS does not cause swamping because $x_{(j-1)}$ being outlying has no influence on the test for its smaller neighbour $x_{(j)}$. However, the limitation of this statistic is that it loses effectiveness when outliers are clustered together.
   
The \emph{sum-robust-sum} (SRS) test statistic for $r$ upper outliers,
  \begin{equation} 
 T_{r,m}^{SRS}= \frac{ \sum_{i=1}^{r}x_{(i)} }{ \sum_{i=m+1}^{n} x_{(i)} },~~m\geq 1~,
 \label{eq:SRS}
  \end{equation}
is a modification of another classic test statistic \citep{Lewis1979,Chikka1983}, here referred to as \emph{sum-sum} (SS) statistic. $m$ is again a pre-specified maximal number of outliers, and when $m=0$, the SS statistic is recovered. In the classical form ($m=0$), the test is equivalent to a likelihood ratio test when the outliers also come from an exponential \citep{Bala}.  Due to the sum over $r$ in the numerator, SRS/SS suffers from swamping. Nevertheless, it is not susceptible to masking because it uses the observation magnitude rather than differences; i.e., it does not compare $x_{(1)}$ versus $x_{(2)}$, which may be close to each other, but far from the rest of the sample. These test are particularly effective when the outliers are clustered. 
  
Another classic test statistic for $r$ upper outliers is the \emph{Dixon} (D) statistic \citep{Dixon1950}, 
  \begin{equation} 
 T_{r}^{D}= \frac{ x_{(1)} }{ x_{(r+1)} }~,
 \label{eq:D}
  \end{equation}
whose distribution function under the null is given by \citep{Likes1966}. In the outward testing case, the joint distribution function was given by \citep{LinBala2014}. It is often regarded as an inferior alternative to the SS  as it considers only a limited number of points in the dataset. This statistic is less robust and also susceptible to masking compared to the SS.
 
We also include a test from the literature of complex systems on detecting `Dragon King' (DK) outliers \citep{DKpisarenko11}. The statistic for $r$ upper outliers,
  \begin{equation}
  T_{r}^{DK}= \frac{ \sum_{i=1}^{r}z_{i} }{ \sum_{i=r+1}^{n} z_{i} }~\sim F_{2r,2(n-r)},
  \label{eq:DK}
  \end{equation}
uses the weighted spacings $z_{i}=i(x_{(i)}-x_{(i+1)}),~i=1,...,n-1$, $z_n= n x_{(n)}$. The statistic follows an F-distribution under the exponential null distribution. It suffers from both masking and swamping, and is less effective in the presence of multiple clustered outliers because it counts spacings rather than absolutes. This statistic is thus mainly advantageous due to the simplicity of its distribution function under the null. 
  
Under the exponential underlying distribution \eqref{eq:Exp}, the distribution functions of all the test statistics above enjoy the pleasant property of being invariant to $\alpha$. This stems from the R\'enyi representation of spacings \citep{Renyi,Bala}, where, for $E_i\overset{\text{iid}}{\sim}\text{Exp}(\alpha)$, the spacings $S_{i}=X_{(i)}-X_{(i-1)}$ are equal in distribution to $(\alpha i)^{-1}E_i$ where $E_i\overset{\text{iid}}{\sim}\text{Exp}(1)$. Consequently, $\alpha$ cancels out in the ratios of sums of spacings or order statistics (which are themselves sum of spacings). This property is particularly valuable as it avoids the need to estimate $\alpha$ in the presence of outliers.
  
In addition to the test statistics mentioned above, a mixture model is considered as a benchmark,
\begin{equation}
  f(x)= (1-\pi)\alpha \text{exp}\{-\alpha x\} + \pi\phi(x;\mu,\sigma)~,~\alpha,\sigma>0~,
  \label{eq:Mix}
\end{equation}
where the Gaussian density $\phi(x;\mu,\sigma)$ accounts for the outlier regime, and $0\leq\pi\leq1$ is the weight. It is common and natural to consider Gaussian distributions to model outliers in the mixture model \cite{verdinelli,aitkin,hodge}. It is in particular beneficial to use the mixture model when numerous outliers together form a distinct, well-defined distribution of their own. The classification of points as either outliers or not is based on the relative weights of the components. The model parameters are estimated using the EM (Expectation-Maximization) algorithm \citep{Redn}, and a likelihood ratio test against the null ($\pi=0$) is used to generate p-values and estimate the number of outliers ($n\hat\pi$). The major advantage of this approach is that it does not require sequential testing, thereby naturally avoiding masking and swamping. Moreover, the model can be extended beyond the exponential, e.g., to a Weibull or gamma distribution function, without complicating the procedure. It is important to note that this method does not distinguish between inliers and outliers -- i.e., the density $\phi$ can be significant both within and beyond where the null distribution function has substantial mass.

\subsection{Distribution function of test statistics}\label{sec:Derivation}

Let ${{X}_{i}},\ i=1,\ldots ,n$ be i.i.d realizations of a random variable $X$ with exponential distribution
\begin{equation}
 F\left( x \right)=1-{{e}^{-\alpha x}},\quad x\ge 0.
\end{equation}

Define the order statistics as the ordered sample ${{X}_{\left( 1 \right)}}>{{X}_{\left( 2 \right)}}>\cdots >{{X}_{\left( n \right)}}$, and the trimmed sum ${{S}_{m,n}}={{X}_{\left( m+1 \right)}}+\cdots +{{X}_{\left( n \right)}}$, $0\le m<n$, hence the test statistic ${{T}_{j,m}}={{X}_{\left( j \right)}}/{{S}_{m,n}}, j=1,\ldots ,m$. Also define the spacing ${{D}_{j,m}}={{X}_{\left( j \right)}}-{{X}_{\left( m \right)}}$, and use the triple $\left( {{X}_{\left( m \right)}},{{D}_{j,m}},{{S}_{m,n}} \right)$ to derive the distribution of ${{T}_{j,m}}$.

First, we derive the distribution of ${{D}_{j,m}}$ and the joint distribution of $\left( {{X}_{\left( m \right)}}, {{S}_{m,n}} \right)$

\begin{equation} \label{eq:spacing}
f_{{{D}_{j,m}}}\left( d \right)\propto {{\left( 1-{{e}^{-\alpha d}} \right)}^{m-j-1}}{{e}^{-j\alpha d}},\quad d\ge 0
\end{equation}
and 
\begin{equation} \label{eq:lowersum}
f_{\left( {{X}_{\left( m \right)}}, {{S}_{m,n}} \right)}\left( x,s \right)\propto {{e}^{-\alpha \left( s+mx \right)}}\sum\limits_{k=0}^{n-m-1}{{{\left( -1 \right)}^{k}}\binom{n-m}{k}}\left( s-kx \right)_{+}^{n-m-1},\quad 0\le s\le \left( n-m \right)x.
\end{equation}

The joint distribution of the triple is then derived as the product of (\ref{eq:spacing}) and (\ref{eq:lowersum}), presented by
\begin{equation} \label{eq:triple}
 \begin{split}
  f_{\left( {{X}_{\left( m \right)}},{{D}_{j,m}},{{S}_{m,n}} \right)}\left( x,d,s \right)
   &= f_{{{D}_{j,m}}}\left( d \right) f_{\left( {{X}_{\left( m \right)}}, {{S}_{m,n}} \right)}\left( x, s \right) \\
   &\propto {{\left( 1-{{e}^{-\alpha d}} \right)}^{m-j-1}}{{e}^{-\alpha \left( jd+s+mx \right)}}\sum\limits_{k=0}^{n-m-1}{{{\left( -1 \right)}^{k}}\binom{n-m}{k}\left( s-kx \right)_{+}^{n-m-1}}
 \end{split}
\end{equation}
over the region $0\le s\le \left( n-m \right)x,\ d\ge 0$.

Next, apply the change of variable $z=x+d$ and derive $f_{\left( {{X}_{\left( j \right)}},{{D}_{j,m}},{{S}_{m,n}} \right)}\left( z=x+d, d, s \right)$ from (\ref{eq:triple}), then marginalize over $d$ to obtain $f_{\left( {{X}_{\left( j \right)}}, {{S}_{m,n}} \right)}\left( z, s \right)$, which reads
\begin{align} \label{eq:lowersum2}
f_{\left( {{X}_{\left( j \right)}}, {{S}_{m,n}} \right)}\left( z, s \right) \propto 
&\sum_{i=1}^{m-j} (-1)^{m-j-i} \binom{m-j-1}{i-1} e^{-\alpha(s+m z)} \left\{ \sum_{k=1}^{n-m-1} (-1)^k \binom{n-m}{k} k^{n-m-1} \right. \notag\\
& \left. \times \frac{(n-m-1)!}{(-\alpha i)^{n-m}} (A_0+B_0-C_0) + \frac{s^{n-m-1}}{\alpha i} \left(e^{\alpha i (z-\frac{s}{n-m})}-1 \right) \right\}, \quad s\le \left( n-m \right)z
\end{align}
where 
\begin{equation}
A_0 = e^{\alpha i \left(z-\frac{s}{k}\right)_{+}},
\end{equation}
\begin{equation}
B_0 = \sum_{l=1}^{n-m-1}(-1)^l\frac{\left(\alpha i\left(\frac{s}{k}-z\right)_{+}\right)^l}{l!},
\end{equation}
\begin{equation}
C_0 = e^{\alpha i \left(z-\frac{s}{n-m}\right)} \sum_{l=0}^{n-m-1} (-1)^l \frac{\left(\alpha i \left(\frac{s}{k}-\frac{s}{n-m}\right)\right)^l}{l!}.
\end{equation}

Proceed with another change of variable $t=z/s$ to derive $f_{\left( {{T}_{j,m}}, {{S}_{m,n}} \right)}\left( t=z/s, s \right)$, then integrate out $s$ to get the null distribution function of the MRS (for $j\le m-1$),
\begin{align}
f_{T_{j,m}}(t) \propto 
&\sum_{i=1}^{m-j} (-1)^{m-j-i} \binom{m-j-1}{i-1} \left\{ \sum_{k=1}^{n-m-1} (-1)^k \binom{n-m}{k} k^{n-m-1}\frac{A_1+B_1-C_1}{(-i)^{n-m}} +  \notag \right.\\
& \left. \frac{n-m}{i} \left( \frac{1}{\left(1 + i \left(\frac{1}{n-m} - t\right) + m t\right)^{n-m+1}} - \frac{1}{(1 + m t)^{n-m+1}} \right) \right\}, \quad t \ge \frac{1}{n-m}
\end{align}
where 
\begin{equation}
A_1 = \frac{1}{(1 + m t - i (t - \frac{1}{k})_{+})^2},
\end{equation}
\begin{equation}
B_1 = \sum_{l=1}^{n-m-1} (-1)^l(1+l) \frac{\left(i \left(\frac{1}{k} - t\right)_{+}\right)^l}{(1 + m t)^{l+2}},
\end{equation}
\begin{equation}
C_1 = \sum_{l=0}^{n-m-1} (-1)^l(1+l) \frac{\left(i \left(\frac{1}{k} - \frac{1}{n-m}\right)\right)^l}{\left(1 + i \left(\frac{1}{n-m} - t\right) + m t\right)^{l+2}}.
\end{equation}

When $j=m$, the null distribution function of the MRS reads
\begin{equation}
f_{T_{m,m}}(t) \propto 
\sum_{k=0}^{n-m-1} (-1)^k \binom{n-m}{k} \frac{(1-kt)_{+}^{n-m-1}}{(1+mt)^{n-m-1}}
, \quad t \ge \frac{1}{n-m}.
\end{equation}

Next, in order to derive the distribution function of $T_{r,m}=\frac{ \sum_{i=1}^{r}x_{(i)} }{ \sum_{i=m+1}^{n} x_{(i)} }=\frac{{S_{0,r}}}{{{S}_{m,n}}}, r=1,\ldots ,m$, we are interested in the conditional distribution of $\left({{S}_{0,r}}| {{X}_{\left( r+1 \right)}} \right)$, which reads
\begin{equation} 
f_{\left( {{S}_{0,r}}|{{X}_{\left( r+1 \right)}} \right)}\left( s| x \right)\propto {{e}^{-\alpha \left( s-r x \right)}}\left( s-r x \right)^{r-1},\quad s \ge r x.
\end{equation}

The conditional distribution helps us to derive the joint distribution of the triple $\left( {{X}_{\left( r+1 \right)}},{{S}_{0,r}},{{S}_{m,n}} \right)$ given the joint distribution of $\left( {{X}_{\left( r+1 \right)}}, {{S}_{m,n}} \right)$, which can be conveniently obtained from (\ref{eq:lowersum2}). By performing transformation of variables and marginalization on the distribution function of the triple, one can obtain the null distribution function of the SRS. Both it and the null distribution function of the MRS, which is detailed in the paper, are provided as Matlab code and are available for access on GitHub (\texttt{github.com/ranwei-ethz/dfs-appendix}) for direct implementation and verification of the procedures discussed.

\section{Outlier Test Performance}

\subsection{Set-up of synthetic tests and issues}\label{sec:Setup}

\begin{figure}[h!]
    \centering
    \includegraphics[width=0.4\textwidth]{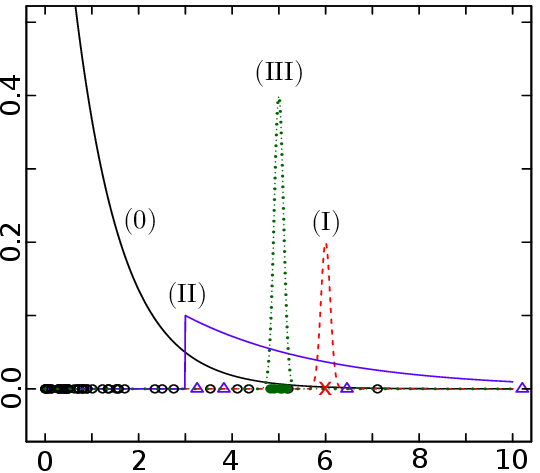}
    \caption{\label{fig:cases} \textbf{Outlier cases}. The null case (0) is the standard exponential for which a realization of 50 points are plotted as open circles. Three outlier cases are considered in addition to the null: (I) a normal distribution function with mean $\mu=6$ and $\sigma=0.1$, presented by a dashed red line, and its single outlier is the red x mark; (II) multiple dispersed outliers $Y_{i}\sim 3+$Exp$(1/\beta),~i=1,...,5$, plotted with a solid blue line for $\beta=4$, with blue triangles indicating (a realization of) the outliers; (III) multiple clustered outliers $Y_{i}\sim$Norm$(\mu,0.1),~i=1,...,5$, plotted with a green dotted line for $\mu=5$, with green dots indicating the outliers. }
\end{figure}

In this section, we compare the performance of the different tests through simulation studies. First, we examine the robustness of test statistics in block tests and contrast them with the mixture test, and study the issues of masking and swamping. Then, we draw comparisons between inward, outward, and mixture tests. Additionally, we examine how misspecification of the null affects test performance. The setup uses a standard exponential sample across four outlier scenarios: (0) no outliers, (I) a single outlier, (II) multiple dispersed outliers, and (III) a cluster of multiple outliers. These scenarios are plotted in Fig.~\ref{fig:cases}.

Let us first revisit the testing methods and the issues of masking and swamping for the sake of clarity.
           
{\bf The block test} examines a fixed set $r$ of suspected outliers in a single test. It determines if all $r$ points are outliers or none. Its success critically depends on accurately choosing $r$, with risks of masking if $r$ is set below the actual number of outliers, and swamping if set above. However, if well specified, it is powerful because of the simultaneous usage of all data.
 
{\bf The inward test} sequentially tests the most extreme data point and removes it if identified as an outlier. The test continues to the next extreme value until a point is not recognized as an outlier. The estimated number of outliers $\hat{k}$ is the number of rejected (marginal) tests. Clearly, this test can suffer from both masking and swamping, and the weaknesses of the inward procedure were cited as motivation for the \emph{outward} test \citep{Rosner1975,Hawkins1980,Kimber1982}:

{\bf The outward test} specifies a maximum number of outliers $r$, and starts by testing if the $r$-th largest value $x_{(r)}$ is an outlier by excluding the larger values $x_{(r-1)}, x_{(r-2)}, ..., x_{(2)}, x_{(1)}$.
 If this test is rejected, then $r$ outliers are identified. If not,  the test progresses to the $(r-1)$-th largest point $x_{(r-1)}$. The test continues until the first detection of outlier. The outward test minimizes the probability and magnitude of both masking and swamping, and has therefore been claimed superior over the inward \citep{Kimber1982,Chikka1983,Balasooriya1994} and received more subsequent development \citep{LinBala2009,LinBala2014}.
  
However, control of the type I error (the probability of a false detection of an outlier) is difficult in the outward test. The test considers the null hypothesis $H_0$ that there are no outliers, with multiple alternatives, $H_j$ that there are $j$ outliers $j=1,...,r$, with test statistic $T_j$. A single rejection of the $r$ tests rejects the null $H_0$. Thus, to achieve an overall type I error level of $0\leq a \leq 1$, e.g., the common level of 0.05 or 0.1, the marginal tests need to have a lower level. The larger $r$ is, the larger the correction will be, and thus the lower the power of the test. This `multiple testing correction' requires knowing the joint and marginal distribution function of, generally dependent, $T_j,~j=1,...,r$. More specifically, one defines all marginal tests to have equal level $b$, i.e.,  $Pr\{ T_j> t_j\}=b,~j=1,...,r$, and the level $b$ is determined such that $Pr\{ T_j\leq t_j ,~j=1,...,r | H_0 \}=1-a$. Clearly $a^r \leq b \leq a$, where the lower bound corresponds to the case of independent tests (the Bonferonni bound), and the upper bound to perfect dependence. For the specific test statistic~\eqref{eq:MRS} discussed below, the joint and marginal distribution functions have been derived as described in \ref{sec:Derivation}.
  
In contrast, for the inward method, the type I error level is equal to the marginal level ($a=b$) because a rejection of the null only happens when the first marginal test (for the largest point, $x_{(1)}$) is rejected. This is a major advantage over the outward procedure in terms of computation and also because no power is lost due to a multiple testing correction.

 \subsection{Performance of block tests}\label{sec:blockpower}
 
Here, we examine the power (at level 0.1) of the range of test statistics employed in block tests, where the block size $r$, and the robustness value $m$ are set to the true number of outliers $k$. We consider three scenarios involving different types of outliers: (I) $n=20$, $k=1$, $X_i\sim$Exp(1), $i=1,...,19$, $X_{20}\sim$Norm$(\mu,0.1)$, $\mu=3,...,10$; (II) $n=50$, $k=5$, $X_{i}\sim 3+$Exp$(1/\beta),~i=46,...,50$, $\beta=1,2,...,6$; (III) $n=50$, $k=5$, $X_i\sim$Exp(1), $i=1,2,...,45$, $X_{i}\sim$Norm$(\mu,0.1),~i=46,...,50$, $\mu=3,4,...,10$. The mixture model~\eqref{eq:Mix} is only estimated in the cases with multiple outliers. 
  
\begin{figure}[h!]
\centering{\includegraphics[width=\textwidth]{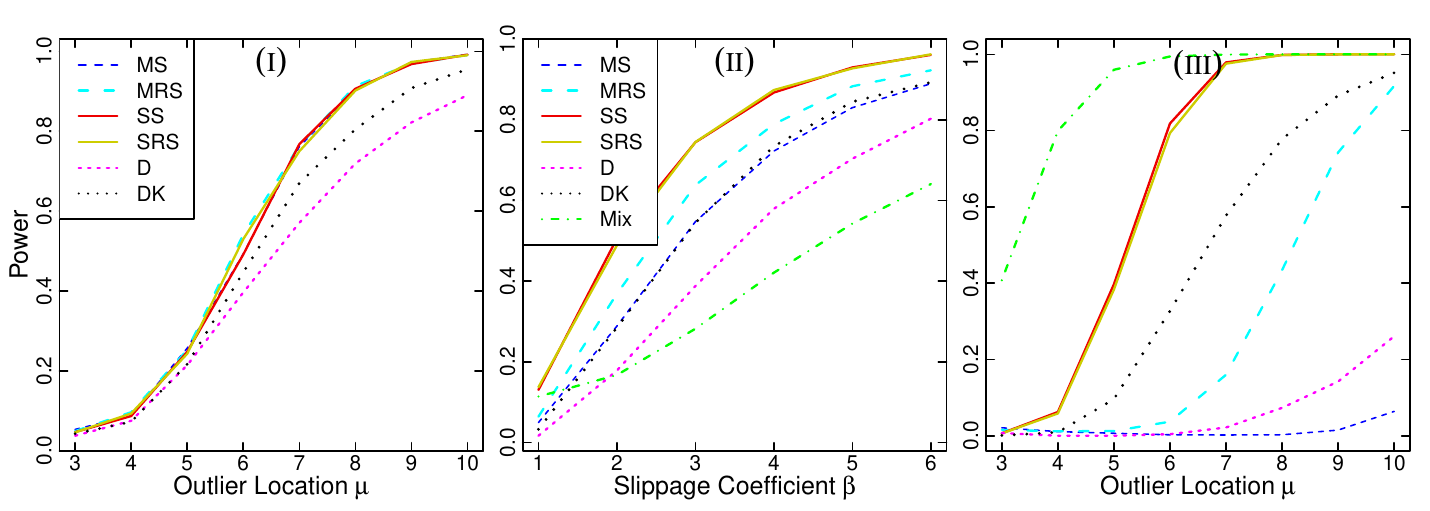}}
  \caption{   Power curves (at level 0.1) for the MS/MRS (\ref{eq:MRS}), SS/SRS (\ref{eq:SRS}), D (\ref{eq:D}), DK (\ref{eq:DK}) statistics, employed in block tests, with the mixture model (\ref{eq:Mix}) estimated in the cases with multiple outliers. The power curves are plotted against the outlier location $\mu$ and the slippage coefficient $\beta$, which are specified in the three outlier cases: (I) $n=20$, $k=1$, $X_i\sim$Exp(1), $i=1,...,19$, $X_{20}\sim$Norm$(\mu,0.1)$; (II) $n=50$, $k=5$, $X_{i}\sim 3+$Exp$(1/\beta),~i=46,...,50$; (III) $n=50$, $k=5$, $X_{i}\sim$Norm$(\mu,0.1),~i=46,...,50$.
}
  \label{fig:PowerTest}
\end{figure}

Fig.~\ref{fig:PowerTest} shows the power curves of the test statistics, for a range of outlier parameters being computed over 10'000 independent simulations. For a single outlier (case I), most of the tests are exactly identical (by definition), with the exception of the DK and D tests, which are weaker. For multiple dispersed outliers (case II), the SS and the SRS tests outperform others and are equivalently effective, followed by the MRS test. The mixture is poorly specified and is thus weakest. For clustered outliers (case III), the performance of the tests varies greatly. Indeed, the test statistics with the sum in the numerator often identifies the cluster of outliers. However, the well specified mixture model is most powerful, also identifying the `outliers' when they are not really outlying but rather a contamination well within the sample (i.e., `inliers'). In detecting multiple outliers, the SS and SRS tests perform almost identically, largely because they both operate under optimal conditions where the block size $r$ and the robustness value $m$ are set to the true number of outliers $k$. The robustness of the SRS statistic is further demonstrated subsequently in context involving masking and swamping.
    
We now present simulation studies to expose the degree to which the different test statistics suffer from masking and swamping in block tests -- that is, how accurately they estimate the number of outliers. This is done by performing the tests on synthetic data for a range of block sizes. The three scenarios considered are: (I) swamping due to a single outlier, $n=30$, $k=1$, $X_i\sim$Exp(1), $i=1,...,29$, $X_{30}\sim$Norm$(8,0.1)$; (II) swamping without masking due to dispersed outliers, $n=30$, $k=5$, $X_i\sim$Exp(1), $i=1,...,25$, $X_{i}\sim 3+$Exp$(1/5),~i=26,...,30$; and (III) swamping with masking due to clustered outliers, $n=30$, $k=5$, $X_i\sim$Exp(1), $i=1,...,25$, $X_{i}\sim $Norm$(8,0.1),~i=26,...,30$. 
  \begin{figure}[h!] 
  \centering{\includegraphics[width=\textwidth]{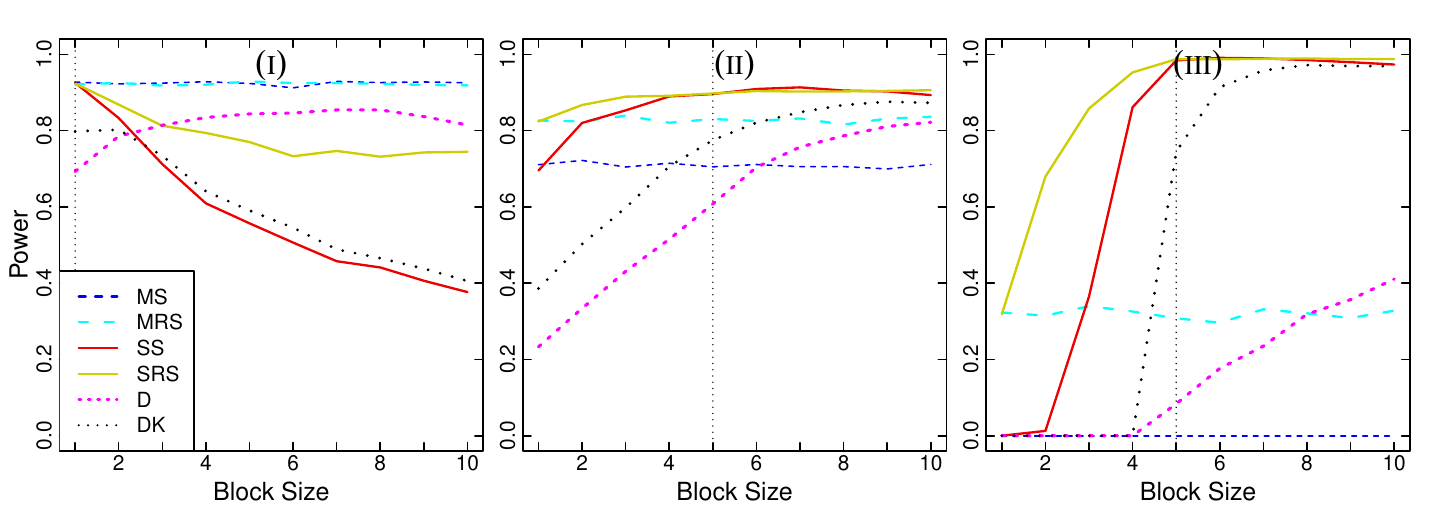}}
  \caption{ Power curves (at level 0.1) for the MS/MRS (\ref{eq:MRS}), SS/SRS (\ref{eq:SRS}), D (\ref{eq:D}), DK (\ref{eq:DK}) statistics, with different block sizes. The data generating processes for the cases are: (I) $n=30$, $k=1$, $X_i\sim$Exp(1), $i=1,...,29$, $X_{30}\sim$Norm$(8,0.1)$; (II) $n=30$, $k=5$, $X_{i}\sim 3+$Exp$(1/5),~i=26,...,30$; and (III) $n=30$, $k=5$, $X_{i}\sim $Norm$(8,0.1),~i=26,...,30$. In each case, the true number of outliers is given by the vertical dotted line.
  }
   \label{fig:MaskSwamp}
  \end{figure}  

Our simulation study determines the frequency at which the tests are rejected, at level 0.1, in 10'000 independent samples, for a range of block sizes ($b=1,2,...,10$). The results are in Fig.~\ref{fig:MaskSwamp}. The MS and MRS tests are not affected by block size since the maximum is always the largest point. In the next section, the inward test will apply the MRS statistic to the largest point, then the second largest, and so on. In that case, the MRS will not cause swamping. As anticipated, masking is problematic for the MS statistic, especially when large observations are densely clustered. Further, as intended, the MRS suffers from masking less than the MS. The SS and SRS tests suffer less from masking and swamping than those based on spacings and maxima. Swamping is pervasive in block testing, even when there is only a single large outlier. That the rejection rate decays slowly as the block size surpasses the true block size indicates that the minimal p-value in the sequence of estimates will not reliably indicate the true block size. These problems motivate sequential testing.

\subsection{Performance of sequential tests}\label{sec:sequential}
        
Here, inward and outward sequential procedures are compared, along with the mixture test. Again the four outlier scenarios visualized in Fig.~\ref{fig:cases} are considered. The tests used are: (i) the outward test with MS, MRS, SS, and SRS statistics; (ii) the inward test with only the MRS statistic, which is necessary to avoid masking and swamping; (iii) the mixture model~\eqref{eq:Mix}; and (iv) the SRS block test, given the correct number of outliers. This last option, which was the best performing block test in Fig.~\ref{fig:PowerTest}, provides a benchmark. 
   
The distribution functions for the test statistics were simulated with 50'000 samples from the null model. All tests were done with a level of 0.1. For the outward test, the level of the marginal tests $b$ was lowered to obtain the overall level of $a=0.1$. For each test, this was done by applying the test on 10'000 independent samples generated from the null, for multiple values of $b$, and selecting $b$ such that $a(b)=0.1\pm0.005$. The resultant marginal levels are in Table~\ref{tab:blevels}. Note how large of an adjustment is needed in the outward test, whereas in the inward test there is no adjustment: $b^{\text{Inward}}=a=0.1$.
      \begin{table}[!h] 
	\begin{center}
	\scalebox{0.9}{
	\begin{tabular}{c  c | c c c c c c c}
	\toprule
	n & r 	& MS & SS & MRS & SRS \\
	50 & 10 & 0.018 & 0.05 	& 0.025 & 0.049 \\
	30 & 5 	& 0.028 & 0.055 & 0.0345 & 0.0575 \\
	15 & 5 	& 0.025 & 0.06 	& 0.036 & 0.056 \\
         \bottomrule
	\end{tabular}	
	}
	\end{center}
		\caption{ Marginal levels (b) for outward tests for different sample sizes (n), maximal number of outliers (r), and robustness value ($m=r$) to obtain an overall type I error level of $a=0.1$ }
	\label{tab:blevels}
\end{table}
  
The results, for slightly different specifications of the four cases, and in order of decreasing sample size, are in Tables~\ref{tab:SeqSim50}, \ref{tab:SeqSim30}, and \ref{tab:SeqSim15}. In case (0), where there are no outliers, the inward and mixture procedures have false positive events that estimate a small number of outliers, whereas the outward procedures falsely identify large numbers of outliers. In case (I) of the sequential procedures, the inward test is most powerful at identifying the single outlier, even matching the power of the block test. The outward tests are substantially weakened, even with relatively small $m=5$. The inward test provides superior estimation of outliers, whereas the other tests tend to overestimate. In case (II), with a cluster of outliers, both the benchmark (the block test) and the inward test perform poorly. They are outperformed by the outward test, which is less susceptible to masking, by design. However, here the mixture approach is both the most powerful and accurate in estimating outlier numeracy. In case (III), with multiple dispersed outliers, all of the inward and outward approaches are similarly competitive, while being slightly dominated by the block test. The mixture approach is weak since the outlier component is poorly specified. For the outward procedure, the MS/MRS statistic dominates the SS statistic.
   
In summary, the inward procedure with the MRS test statistic is more computationally convenient than the outward procedure, commits less severe false positives, and can even be more powerful when identifying single or multiple dispersed outliers. In the event of a dense cluster of outliers, a mixture approach can be more computationally convenient and powerful than the outward approach. Within the outward approach, the MS/MRS statistic is found superior to the SS/SRS statistic, and robust modifications performed similarly.
      
   \begin{table}[!h] 
	\begin{center}
	\scalebox{0.85}{
	\begin{tabular}{c c | c c c c c c c}
	\toprule
  Case 	& Quantity	& 	MS Out & 	SS Out & 	MRS Out & 	SRS Out & 	MRS In & 	Mix 	& SRS Block	\\
    (0)	&Rej. Rate	&	0.11	& 0.10		&	 0.11	& 0.10		& 0.10		& 0.14		&	0.10	\\
    (0)	&$\widehat{k}$	&(3,6,9)	& (5,9,10)	& (3,6,9)	& (5,9,10)	& (1,1,3)	& (2,2,4)	&		\\
   (I) &Rej. Rate	& 0.30 		&0.22		& 0.30		& 0.22		& 0.64		& 0.09		&	0.69	\\
   (I) &$\widehat{k}$	&(2,3,6)	&(2,5,10)	&(2,3,7)	&(2,5,10)	&(1,1,2)	&(2,2,2)	&	$=1$	\\
   (II)&Rej. Rate	&0.91 		&0.75		& 0.89		& 0.75		& 0.04		& 0.95		&	0.38	\\
   (II)&$\widehat{k}$	&(5,7,8)	&(5,7,10)	&(5,7,9)	&(5,7,10)	&(1,9,10)	&(5,5,6)	&	$=5$	\\
   (III) &Rej. Rate	&0.96 		&0.96		& 0.97		& 0.96		& 0.95		& 0.63		&	0.98	\\
   (III) &$\widehat{k}$	&(5,6,8)	&(4,6,10)	&(5,6,9)	&(4,6,10)	&(6,7,10)	&(3,10,10)	&	$=5$	\\
         \bottomrule
	\end{tabular}	
	}
	\end{center}
		\caption{ $n=50$ (sample size), $m=10$ (robustness value). Summary of tests over 5000 repeated simulations of four cases: (0) the null case ($X\sim$Exp(1)), (I) a single large outlier ($X_i\sim$Exp(1), $i=1,...,49$;  $X_{50}\sim$Norm$(7,0.1)$), (II) a cluster of multiple outliers ($X_i\sim$Exp(1), $i=1,...,45$;  $X_{i}\sim$Norm$(5,0.1),~i=46,...,50$); (III) multiple dispersed outliers ($X_i\sim$Exp(1), $i=1,...,45$;  $X_{i}\sim$ max$(\{X_i:i=1,...,45\}) $+Exp$(1/5)$,~$i=46,...,50$). The rejection rate and  the median $\widehat{k}$ and quartiles of the estimated number of outliers (in the event of a rejection) are given in alternating rows. }
	\label{tab:SeqSim50}
\end{table}
   \begin{table}[!h] 
	\begin{center}
	\scalebox{0.85}{
	\begin{tabular}{c c | c c c c c c c}
	\toprule
  Case 	& Quantity	& 	MS Out & 	SS Out & 	MRS Out & 	SRS Out & 	MRS In & Mix 		& SRS Block	\\
    (0)	&Rej. Rate	&	0.11	& 0.11		&	 0.11	& 0.11		& 0.11		& 0.16		&	0.10	\\
    (0)	&$\widehat{k}$	&(2,3,5)	& (4,5,5)	& (2,4,5)	& (3,5,5)	& (1,1,3)	& (2,2,5)	&		\\
   (I) &Rej. Rate	& 0.45 		&0.32		& 0.43		& 0.33		& 0.72		& 0.08		&	0.75	\\
   (I) &$\widehat{k}$	&(1,2,3)	&(1,3,5)	&(1,2,3)	&(1,2,5)	&(1,1,2)	&(2,2,2)	&	$=1$	\\
   (II)&Rej. Rate	&0.72 		&0.63		& 0.73		& 0.64		& 0.08		& 0.96		&	0.36	\\
   (II)&$\widehat{k}$	&(3,4,5)	&(3,4,5)	&(3,4,5)	&(3,4,5)	&(4,5,5)	&(3,3,3)	&	$=3$	\\
   (III) &Rej. Rate	&0.87 		&0.86		& 0.89		& 0.86		& 0.88		& 0.50		&	0.90	\\
   (III) &$\widehat{k}$	&(2,4,4)	&(2,4,5)	&(2,4,5)	&(3,4,5)	&(3,4,5)	&(2,5,7)	&	$=3$	\\
         \bottomrule
	\end{tabular}	
	}
	\end{center}
	\caption{ $n=30$ (sample size), $m=5$ (robustness value). Summary of tests over 5000 repeated simulations of four cases: (0) the null case ($X_i\sim$Exp(1)), (I) a single large outlier ($X_i\sim$Exp(1), $i=1,...,29$; $X_{30}\sim$Norm$(7,0.1)$), (II) a cluster of multiple outliers ($X_i\sim$Exp(1), $i=1,...,27$; $X_{i}\sim$Norm$(5,0.1),~i=28,29,30$), (III) multiple dispersed outliers ($X_i\sim$Exp(1), $i=1,...,27$; $X_{i}\sim$ max$(\{X_i:i=1,...,27\}) +$Exp$(1/5),~i=28,29,30$). The rejection rate and the median $\widehat{k}$ and quartiles of the estimated number of outliers (in the event of a rejection) are given in alternating rows.}
	\label{tab:SeqSim30}
\end{table}
   \begin{table}[!h] 
	\begin{center}
	\scalebox{0.85}{
	\begin{tabular}{c c | c c c c c c c}
	\toprule
  Case 	& Quantity	& 	MS Out & 	SS Out & 	MRS Out & 	SRS Out & 	MRS In & Mix 		& SRS Block	\\
    (0)	&Rej. Rate	&	0.11	& 0.11		&	 0.11	& 0.11		& 0.08		& 0.16		&	0.10	\\
    (0)	&$\widehat{k}$	&(2,3,4)	& (3,5,5)	& (2,3,5)	& (3,5,5)	& (1,2,4)	& (2,3,5)	&		\\
   (I) &Rej. Rate	&  	0.25	&0.22		& 0.23		& 0.20		& 0.30		& 0.14		&	0.30	\\
   (I) &$\widehat{k}$	&(2,3,4)	&(2,4,5)	&(2,3,4)	&(2,4,5)	&(1,2,3)	&(2,2,4)	&	$=1$	\\
   (II)&Rej. Rate	&0.42 		&0.42		& 0.43		& 0.41		& 0.04		& 0.93		&	0.13	\\
   (II)&$\widehat{k}$	&(3,4,5)	&(4,5,5)	&(3,4,5)	&(3,5,5)	&(3,4,5)	&(3,3,3)	&	$=3$	\\
   (III) &Rej. Rate	&0.63 		&0.62		& 0.64		& 0.62		& 0.63		& 0.37		&	0.66	\\
   (III) &$\widehat{k}$	&(2,3,4)	&(2,4,5)	&(2,3,4)	&(2,4,5)	&(2,3,5)	&(2,3,4)	&	$=3$	\\
         \bottomrule
	\end{tabular}	
	}
	\end{center}
		\caption{ $n=15$ (sample size), $m=5$ (robustness value). Summary of tests over 5000 repeated simulations of four cases: (0) the null case ($X_i\sim$Exp(1)), (I) a single large outlier ($X_i\sim$Exp(1), $i=1,...,14$; $X_{15}\sim$Norm$(4,0.1)$), (II) a cluster of multiple outliers ($X_i\sim$Exp(1), $i=1,...,12$; $X_{i}\sim$Norm$(4,0.1),~i=13,14,15$), (III) multiple dispersed outliers ($X_i\sim$Exp(1), $i=1,...,12$, $X_{i}\sim$max$(\{X_i:i=1,...,12\}) +$Exp$(1/5),~i=13,14,15$). The rejection rate and the median $\widehat{k}$ and quartiles of the estimated number of outliers (in the event of a rejection) are given in alternating rows.}
	\label{tab:SeqSim15}
\end{table}
 
 \subsection{Robustness to null mis-specification}\label{sec:robust}
 
\begin{figure}[h!]
  \centering{\includegraphics[width=\textwidth]{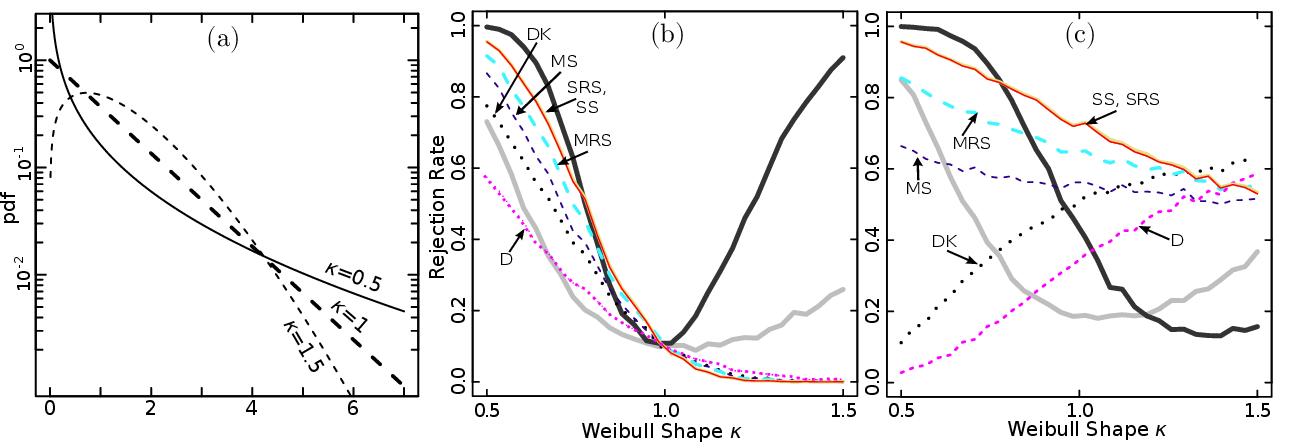}}
  \caption{ \textbf{Test robustness} Panel (a): The Weibull PDF~\eqref{eq:weibull} plotted for parameters $(\kappa,\beta)$ equal to $(0.5,0.4),~(1,1)$ and $(1.5,1.5)$. Panel (b):  The frequency of rejection of the null of no outliers, at level 0.1, in the presense of no outliers, for block tests for $r=3$ outliers, assuming an exponential null model, when the data is generated from a Weibull for a range of shape parameters $\kappa$. Panel (c): The frequency of rejection of the null using a level 0.1, of the block tests for $r=3$ outliers, with the same setup as frame (b), except that 3 outliers are truly present. The models for the cases are: (b) $X_i\sim$Weibull$(\kappa,1)$, $i=1,...,30$; (c) $X_i\sim$Weibull$(\kappa,1)$, $i=1,...,27$,  $X_{i}\sim$max$(\{X_i:i=1,...,27\})+$Exp$(1/3),~i=28,29,30$). For each case, simulation and testing were performed 1000 times for $\kappa$ sweeping 0.5 to 1.5. The tests are colour coded: SS (red solid), SRS with $m=r$ (yellow solid), MS (blue dashed), MRS with $m=r$ (turquoise heavy dashed), D (magenta light dotted), DK (black dotted). In both frames, the black heavy solid line is the power of the likelihood ratio test of the Weibull versus the exponential on the data (including outliers). Similarily the grey heavy solid line is for the Kolmogorov-Smirnov test. }
  \label{fig:Robust}
  \end{figure}

 In practice, the correct specification of the null/main model is of considerable importance. Here, the sensitivity of the rate of false positives (level / type I error), and true positives (power), to the degree of misspecification of the null are exposed via a simulation study, for the battery of test statistics implemented in block tests. We consider simulating data from a Weibull distribution,
 \begin{equation}
  F(x)=1-\text{exp}\{ -(x/\tau)^\kappa \},~x\geq 0,~\tau,\kappa>0~,
  \label{eq:weibull}
 \end{equation}
 which is exponential ($\alpha=\tau^{-1}$) when $\kappa=1$, is fat tailed for $\kappa<1$, and becomes concentrated at $\tau$ as $\kappa$ becomes large. The results of the simulation study are presented in Fig.~\ref{fig:Robust} and can be described as follows.
 
Panel (b) concerns the rate of false positives where $r=3$ outliers are tested, with level $a=0.1$, in a Weibull~\eqref{eq:weibull} sample of size $n=30$, for a range of shape parameters $\kappa$, without outliers. When $\kappa<1$, the distribution function is fat tailed, having many events that are large, and thus the tests falsely identify many points as outliers. This is problematic in practice (with small to moderate sample sizes), because one does not know what the true null model is. For instance, with $n=30$, even when the true distribution function is considered as an alternative model versus the exponential, and using the powerful likelihood ratio test, 50 percent of the time (for $\kappa\approx 0.6$), one will not reject the exponential model at a level of 0.1. In this case, when falsely retaining the exponential model, the type I error will be between 0.3 and 0.5, depending on the selected test statistic. The KS test of compatibility of the data with the exponential distribution function is even less powerful, allowing for more severe false positives. 
 
 Case (c) considers the frequency of true positives (power). The setup is the same as above, but 3 dispersed outliers are included. When the Weibull distribution function becomes less fat tailed, the power of the SRS and MRS tests decreases whereas the power of the D and DK tests increases. Here, with $n=30$, for the tests of the Weibull versus the exponential, including the outliers in the sample, there is a high probability (0.6-0.8) of not rejecting the exponential model when $1<\kappa<1.5$, where the power of some of the tests is weakened.
 
 It is clear that the power, and especially the level, are highly sensitive to the validity of the exponential model, and misspecification of the null can lead to erroneous inference. This has important implications for the practical application of the tests. In particular, one should have a sufficiently large sample to diagnose the validity of the null, and not blindly accept/reject the result of the test and its diagnostics.
 
 \section{Generality of exponential distribution}\label{sec:general}
 
 It is important to note that outlier tests with both the Pareto and exponential underlying distributions are generally applicable to data having approximately Pareto or exponential tails. This follows from the well known Pickands-Balkema-de Haan theorem of Extreme Value Theory (EVT), that states \citep{Embrechts}: For a broad range of distributions, for random variable $X$, with sufficiently high threshold $u$, the excess distribution function, $F_{u}(x)=P\{X-u\leq x | X-u>0\}$ (i.e., the tail of the distribution function), is approximated by the GPD (Generalized Pareto Distribution Function),
 \begin{equation}
 \label{eq:GPD}
 GPD(x;\xi,\beta,\mu)=\begin{cases}
   1-\left( 1-\xi (x-\mu)/\beta \right) ^{-1/\xi}  , & \text{if $\xi \neq 0$}\\
   1 - \text{exp}(-(x-\mu)/\beta), & \text{if $\xi =0$ },
 \end{cases}  
\end{equation}
 in the sense that, 
 \begin{equation}
 \text{lim}_{u \rightarrow  \infty} \text{sup}_{0\leq x } \left|~F_{u}(x)-GPD\left(x|\xi,\beta(u),\mu\right) \right| = 0~,~~~\beta(u)>0, \forall u.
 \label{eq:PBDH}
\end{equation}
 If $\xi=0$ (the Gumbel case), then the GPD~\eqref{eq:GPD} is exponential with lower truncation $\mu=u$ and scale parameter $\beta=1/\alpha$. This case includes common distributions such as the exponential (obviously), the Normal, and even some fat-tailed ones such as the Lognormal. If $\xi>0$ (the Fr\'echet case), the GPD~\eqref{eq:GPD} is (generalized) Pareto with $\mu=u$, $\sigma=u/\alpha$, and $\xi=1/\alpha$. This case includes heavy tailed distributions such as the Pareto and Log-gamma. The only other case ($\xi<0$: the Weibull case) is where the distribution function has a finite upper endpoint, which is of less interest in outlier detection. Therefore, since a Pareto tail can be transformed to an exponential one, outlier testing in exponential samples is (asymptotically) extremely general!
   
  Since the GPD approximation (\ref{eq:PBDH}) is only asymptotically valid, one must select a sufficiently large lower threshold $u$ before applying outlier tests. The problem of threshold selection is a tradeoff between bias and variance, and is the primary statistical issue in the EVT literature, where it is referred to as sample fraction selection. In the physics literature, threshold selection and goodness of fit diagnostics are important for the interpretation of mechanisms underlying power laws found in datasets. There are a variety of tools available for this task.   
   
  The classic `Hill plot' method \cite{hill} for threshold selection consists of estimating the model for a range of thresholds and selecting the lowest threshold (the largest sample fraction) where the estimate is `stable' -- i.e., consistent with values of the estimate for larger thresholds. See Fig.~\ref{fig:case1} for an example. Of course, one can also look for statistically significant changes in the estimated parameter relative to the hopefully stable value obtained deeper in the tail \cite{hill,hall,bauke2007}, however more powerful principled methods exist (see e.g., \cite{beirlant2006statistics,GomesOliviera} for a review). For instance, let us mention the methods based on minimizing the asymptotic mean square error of the estimate. This requires assuming the (class of) distribution beyond the power law tail \cite{hall}, or using bootstrap methods \cite{danielsson,GomesOliviera}. 
  
  These methods have not been extensively adopted outside of the EVT literature. For instance, the most highly cited paper on the estimation of power laws and sample fractions \cite{clauset} does not mention the sample fraction estimation literature. However a subsequent work \cite{virkar2014power}, extending the method to binned/aggregated data, does provide such references. The popular work \cite{clauset} suggests choosing the pair of u and $\alpha$ that have the smallest KSD (Kolmogorov-Smirnov distance). The KSD criterion penalizes error, and rewards sample size. However, as noted by \cite{Corral,Corral2}, comparing KSD across samples of different size is not necessarily consistent as the KSD simply scales with growing sample size like $\sim 1/n^{0.5}$. Further, in \cite{clauset}, no argument was given why this is optimal. In \cite{Corral,Corral2}, it was shown that the method fails when the distribution has a power tail whose parameter changes from one value to another. Originally, \cite{hill,hall} proposed applying a test for decreasing $u$, and selecting $u$ at the value before the first value where the test is rejected. In \cite{Corral}, a similar approach was proposed based on the KS test, where instead one would select the largest sample that could not be rejected, regardless of if rejection occurs at higher thresholds.
  
  These methods can be thought of as outlier tests, where `lower outliers' are points below the tail threshold $u$, that are discordant with the tail. 
  However, instead of elaborating on this, a more general automatic approach is recommended: One should fit both the exponential, and a more complicated density to the range of upper samples, and identify the threshold at which the complicated density is not significantly better. If the more complicated density is sufficiently flexible, this should determine that the exponential provides a good approximation and is sufficient to describe the data above the threshold. One could consider comparing nested models with the likelihood ratio test, however this is only a comparison with a specific alternative model. For a more general alternative model, one can use a non-parametric estimator, such as the logspline estimator (available in R:locfit) \cite{kooperberg}. One can then compare the null with this alternative with the Akaike Information Criterion (AIC). In the presense of clear outliers, one may wish to use estimators that censor, or are robust to the outliers.
  
  Concerning outlier testing, it is useful to estimate the sample fraction to have an idea of where the tail approximation begins to apply. However, tests can often accept a model for a larger sample but reject it in the tail! Thus, one should apply outlier tests for a range of lower thresholds and look for stability in outlier test results for data that do not violate the null. That is, letting $n_u \leq n$ be the size of the largest upper sample that can be defended based on the methods discussed above, an outlier test should be applied to the upper samples consisting of the $n_u, n_{u-1},...,10+r$ largest points, where r is the expected number of outliers, and where one should certainly not consider samples of size smaller than ten. Consistent identification of outliers in these upper subsamples, where the GPD approximation \eqref{eq:PBDH} is most relevant, and where the null model cannot be rejected, should be interpreted as a robust result. This algorithm involves $c=n_u-(10+r)$ consecutive dependent tests, which gives multiple chances for a false positive. However, under the null, the probability of rejecting $c>1$ consecutive tests, decreases as c increases. Based on simulation studies with the range of models considered within this work, we offer as a rough rule of thumb, that for a sample of size $10<n<100$, one should require a run of $c=n/10$ tests to be rejected to maintain control of the type I error.

\section{Case study and `Dragon Kings'}\label{sec:case}

\subsection{Pareto distributions and beyond: the Dragon-King hypothesis}

  Outlier detection with exponential underlying distribution has been primarily motivated by reliability engineering applications. Switching perspective from reliability to risk, the exponential of an exponential variable has the (heavy-tailed) Pareto distribution \eqref{eq:Pareto} that is typically used for modeling extremes in both natural and social sciences: earthquake energies, stock prices, claims in non-life insurance, etc. \citep{Embrechts,Mitzenmacher04,NewmanMEJ05,Sornette2006}. 
  
  The Pareto distribution is unique in that it is scale invariant \citep{DubrulleGranerSor98,LesneLaq11}, 
  suggesting that events of all sizes -- including extremely large ones -- are generated by a single mechanism operating at different scales. This feature allows this single parsimonious distribution function to generate a broad range of event sizes. Thus, if a phenomenon is scale invariant, then extreme events are not predictable and there is nothing anomalous about them as there is nothing to distinguish these events from their smaller siblings, other than their resultant size. This reasoning has been advanced to explain the extreme difficulties in forecasting large earthquakes \citep{Gelleretal1997}: according to the approximate scale invariance of the Gutenberg-Richter law, large earthquakes are just earthquakes that started small... and did not stop growing.
 
  However, a number of studies have found either strong or, in other cases, suggestive evidence that there are extreme events `beyond' the Pareto sample \citep{SornetteDK2009,SorOuiDrag}, i.e., outliers, inspiring the concept of the `Dragon King' (DK) \citep{SornetteDK2009} event. DK embody a double metaphor implying that an event is both extremely large (a king \citep{LaherSornette}), and generated from a unique mechanism/origin (a dragon) relative to other events in the system/sample. The hypothesis advanced in \citep{SornetteDK2009,SorOuiDrag} is that DK events are generated by a distinct mechanism (e.g., positive feedback) that intermittently amplifies extreme events, leading to the generation of runaway disasters as well as extraordinary opportunities/successes. 
  Due to the uniqueness of such events, there is hope that such extremes may exhibit precursory signs, disclosing some predictability. The identification of the existence of such phenomena is also clearly important -- for example, with applications in risk management. Examples of such DK events have been proposed to include failures of material systems, landslides \citep{Lei2023} and some large earthquakes in geophysics, financial crashes in economics \citep{Johansen1998Crashes,Johansen2001Drawdowns,Filimonov2015}, and epileptic seizures and human parturition in biology  \citep{SornetteDK2009,SorOuiDrag}.
  Identifying DKs with convincing statistical significance is a prerequisite to the investigation of their origin, understanding their generating mechanisms, and developing forecasting methods, controls, and resilient system designs. Motivated by these considerations, and to provide pedagogical examples, five case studies are considered where DK events are tested as statistical outliers. The case study on financial market crashes is given below, and the other four in the Appendix.

\subsection{Financial crashes}
  
  It is well known that crashes in financial markets occur frequently and can have a significant effect not only on market participants, but also on the broader economy. It is often thought that financial markets are unpredictable -- i.e., they are scale invariant / fractal ~\citep{Mandelbrot,SornetteWhyMarketsCrash} (Pareto distributed). However, in \citep{Johansen2001Drawdowns,Johansen1998Crashes,Filimonov2015} it was found that the sample of crash sizes -- measured from the peak to the valley of the event (so-called drawdowns) -- contained outliers (defined below). However, the statistical test used in \citep{Filimonov2015} contains an error in the distribution function of the marginal test statistics, and \citep{Johansen1998Crashes,Johansen2001Drawdowns} did not use standard outlier tests. To correct this, and provide an example, this problem is revisited with the same data. The data are the drawdowns computed for the eleven most actively traded Futures Contracts on the American and European Indices\footnote{
  US: 1) ES, S\&P 500, E-mini; 2) NQ, NASDAQ, E-mini ; 3) DJ, Dow Jones, E-mini. 
  European: 4) AEX, Netherlands; 5) CAC, CAC40, France; 6) DAX, Germany; 7) FTSE, UK; 8) IBEX, Spain; 9) OMX, OMX Stockholm 30, Sweden; 10) SMI, Switzerland; 11) STOXX, Euro STOXX, Europe. }, from January 1, 2005 to December 30, 2011. 
  
   A peak-to-valley measure of the size of intra-day financial crashes is considered: an \emph{$\epsilon$-drawdown} (hereforth referred to simply as a drawdown) is the total cumulative return of a negative run in price over time, with some specified tolerance for small positive changes along the way \citep{Johansen2001Drawdowns}. A \emph{drawup} is its positive counterpart. This is an interesting measure of risk because it captures the transient dependence of price changes in time, whereas studying the unconditional distribution of returns does not. More specifically, considering one trading day $\left[ t_0,t_1 \right]$, prices taken at intervals of width $\Delta$ are $p_i=p(t_0+i\Delta)$, $i=1,...,n=\lfloor (t_1-t_0) / \Delta \rfloor$. The \emph{returns} are then $r_{i}=\text{log}(p_i/p_{i-1})$. One starts at the first negative return $i_0 = \text{min}\{ i : r_i < 0 \}$. Then, the cumulative return,
  \begin{equation}
    r_{i_0,i}=\sum_{j=i_0}^{i} r_j = \text{log}(p_{i}/p_{i_0}),~i>i_0~,
    \label{eq:ddreturn}
    \nonumber
  \end{equation}
  tracks the negative growth of the drawdown, continuing for $i=i_0,i_0+1\dots$ until the first value of $i$, say $i_2$, such that the cumulative return has appeared to reverse direction, relative to its lowest point:
  \begin{equation}
  r_{i_0,i_2} - \text{min}_{ i_0 \leq j \leq i_2 } r_{i_0,j} > \epsilon \sigma~.
      \label{eq:break}
      \nonumber
  \end{equation}
Parameter $\epsilon \geq 0$ tunes the tolerance of moves in the opposite direction, and $\sigma$ is the standard deviation of the returns from the previous trading day. The inclusion of $\sigma$ makes the tolerance adaptive, which allows for volatility regimes. Finally, stepping backwards from $i_2$, which is the index of a positive change, the drawdown is defined to have occurred from the start $i_0$ to the lowest point, which occurs at $i_1=\text{argmin}_{j\in(i_0 \leq j \leq i_2)} r_{i_0,j}$. From the next index, $i_1+1$, a drawup is defined to begin and computed in a similar way. Drawdowns and drawups alternate in this contiguous way, for the entire trading day.
  
\begin{figure}[h!] 
  \centering{\includegraphics[width=\textwidth]{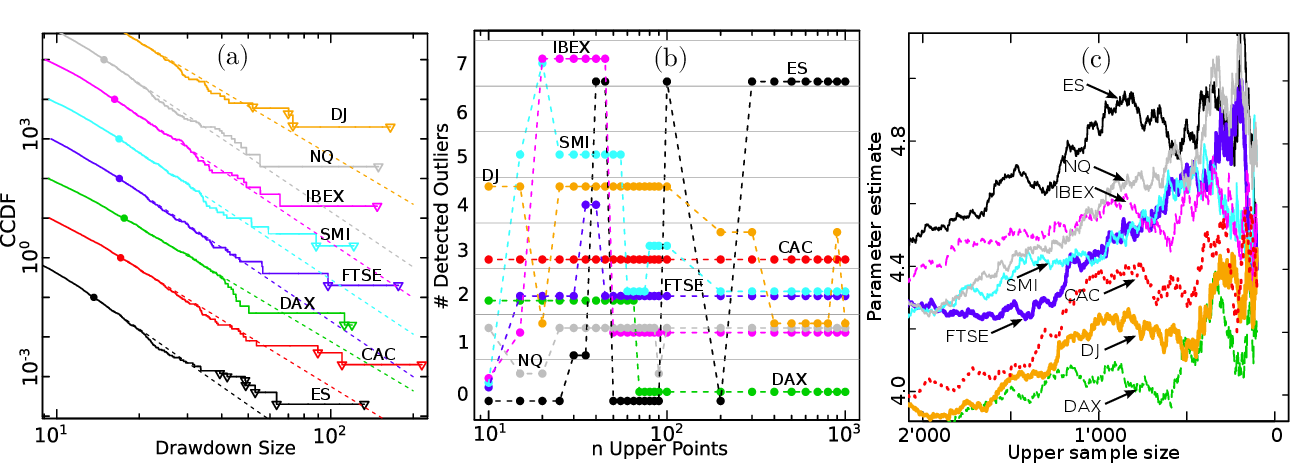}}
  \caption{\textbf{Financial Market Crashes.} \textbf{(a):} The 5000 largest drawdowns for each of the 8 futures contracts thought to contain outliers, plotted according to their empirical CCDF in double logarithmic scale. For clarity, each CCDF above the black one is multiplied by 10 relative to the one beneath it. The Pareto distribution function with MLE parameter for the top 500 points is given by the dashed lines, starting at the solid dot. The triangles identify the points that were identified as outliers based on the interpretation of panel (b). \textbf{(b)} The number of identified outliers is plotted against sample size where the MRS test~\eqref{eq:MRS} with level $a=0.1$ has been applied inward with $m=r=10$, for a range of sample sizes $n$, for each contract in (a) with the same colour coding. \textbf{(c)} Hill plot: The estimated tail exponent is plotted for a range of upper sample sizes.
  (see online version for colour)}
  \label{fig:case1}
  \end{figure}

  In panel (a) of Fig.~\ref{fig:case1}, for the eight contracts thought to contain an outlier, the largest 5000 drawdowns are plotted according to their empirical CCDF (complementary CDF, i.e., $1-F(x)$). The empirical CCDF appear approximately linear in the double logarithmic scale, indicating a qualitatively good fit, with the exception of some outliers. There are also some additional differences in the tail. For instance, the tail of the CCDF drops beneath the Pareto fit before crossing back to form the outlying empirical tail. This could suggest an amplification mechanism operating above a threshold size. In panel (c), the Hill plot is given, where the MLE for the tail exponent $\alpha$ is plotted for a range of upper sample sizes. The parameters tend to have an increasing trend, indicating slight convexity in the CCDF in panel (a), and thus a loss of outlier testing power for large sample sizes (Sec.~\ref{sec:robust}). Based on the Hill plot, the estimator for the top 1'000 points appears to be approximately stable for most of the contracts. For systematic threshold / sample fraction selection, three methods are used: 1. comparing the AIC of exponential and nonparametric (R:logspline) fits, 2. selecting the smallest KSD as recommended in \cite{clauset}, and 3. selecting the smallest threshold where the KS test p-value is above 0.10. The results of these methods are given in Tab.~\ref{tab:Thres}. All but one of the 24 tests select at least the top 1'000 points, thus upper samples of this size and smaller will be considered for outlier testing.
 
      \begin{table}[!h] 
	\begin{center}
	\scalebox{0.9}{
	\begin{tabular}{c | c c c c c c c c}
	\toprule
	Test	&ES 	& CAC& DAX & FTSE 	& SMI & IBEX & NQ & DJ 		\\
	AIC	&1'184 	& 1'214& 2'290 & 2'734 	& 2'704 & 2'055 & 1'154 & 3'757 	\\
	KSD	&1'049 	& 1'115& 1'520 & 3'144 	& 609 & 1'501 & 1'074 & 1'134 	\\
	p	&1'985 	& 1'323& 2'403 & 3'714 	& 2'701 & 2'255 & 3'123 & 4'000 	\\
         \bottomrule
	\end{tabular}	
	}
	\end{center}
		\caption{ The selected number of points in the upper sample based on comparing the AIC of the null and the nonparametric model, minimizing the KSD, and the largest upper sample with KS test p value greater than 0.1.}
	\label{tab:Thres}
\end{table}

  In panel (a) of Fig.~\ref{fig:case1}, the apparent outliers are large and dispersed. Thus, the MRS test statistic \eqref{eq:MRS} should be powerful (Sec.~\ref{sec:sequential}) and can be applied inward for a range of thresholds, requiring a fraction of the computation of outward testing. For each dataset, the inward test was performed -- with MRS, $m=10$, level $a=0.1$, and upper sample size ranging from $n=10$ to $n=1000$. For all contracts, excluding AEX, OMX, and STOXX, at least 1 outlier was found and are indicated in Panel (b) of Fig.~\ref{fig:case1}. For some of the contracts, the results are quite stable across sample size (e.g., CAC and FTSE). For others, the impurity of the distribution function plays a role in the interpretation. For instance, for DAX, two outliers are detected once the test is restricted to the bent-down tail. For ES, choosing between zero and seven outliers is more subjective -- are there multiple outliers, or does the tail grow heavier? For IBEX, it is clear that the identification of seven outliers is due to the dip in the empirical CCDF occurring between drawdown size of twenty and thirty. The alternative choice of 1 outlier is more stable with respect to a broad range of values of $n$. The interpreted outliers are indicated in panel (a).    
  
  The largest outliers coincide with major news events: The 07 July 2005 London bombings coincided with the largest outliers of CAC, DAX, FTSE, SMI, and IBEX -- all being based on European indices. Further, DAX and CAC each have an outlier corresponding to the `Mini Flash Crash' of 27 Dec. 2010 (e.g., see \citep{bundesbank}). All American contracts (ES, DJ, and NQ) have their largest outliers coinciding with the infamous `2010 Flash Crash' of 6 May 2010. We thus observe that outliers occur either due to some exogenous impacts (London bombings) or as a result of an endogenous transiently unstable dynamics (flash crash). Indeed, in \citep{FiliSorReflex,FiliSorBichMay14,Wheatley2019EndoExo,Wehrli2021Hawkes,Wehrli2022FlashCrashes,Wehrli2022VolatilityPuzzle}, it was suggested that financial markets exhibit a significant endogeneity or `reflexivity', in the sense that nowadays up to 70-80\% of trades occurring at the time scales of fractions of seconds to tens of minutes are motivated (or triggered) by previous trades. In this framework \citep{FiliSorReflex,FiliSorBichMay14}, Dragon Kings emerge when the market dynamics become critical and super-critical, that is when the future trades are triggered only by previous trades and not by news, making the financial markets essentially self-referential in these periods. Thus, some of the outliers can be classified as Dragon-King drawdowns.

\clearpage

  \section{Discussion}

  We have provided a comprehensive study of outlier detection in the general case of samples with exponential and Pareto tails. By considering a variety of test statistics and outlier scenarios, many useful insights are made available to practitioners. Further, a simple yet novel modification of classical test statistics was shown to make the convenient inward test competitive with the relatively arduous outward test. 
  
  Insights include that one should select the correct test statistics based on the nature of the suspected outliers. For instance, a mixture model can be very useful for clustered outliers, whereas an inward test with a MS type statistic will be powerless. Next, the power and level of outlier tests are highly sensitive to the correct specification of the null. For robust results, it may be better to focus on the tail of the sample, where EVT ensures that the best approximation is attained. If the approximation is poor even in the tail, then one should choose a better null model to avoid spurious inference. Further, tests should be applied for a range of upper samples (growing lower threshold) and consistent rejection required for a robust rejection to be verified.
  
  In the case studies, the concept of Dragon King events was introduced. This stresses that some outliers are meaningful, and perhaps special. Further, one should certainly not simply discard these outliers but rather focus on understanding them. Significant outliers were found in the sizes of financial returns and crashes, epidemic fatalities, nuclear power generation accidents, and city sizes within countries. In the cases of financial crashes and nuclear accidents, the existence of Dragon Kings should be considered in the assessment of risk.

  \section{Acknowledgement}
  
  We thank Doctor Spencer Wheatley for his contribution to the preliminary version of this paper. His assistance in the initial analysis especially on the case studies was crucial for the work.
~\\  

  D. S. was partially supported by the National Natural Science Foundation of China (Grant No. T2350710802 and No. U2039202), Shenzhen Science and Technology Innovation Commission Project (Grants No. GJHZ20210705141805017 and No. K23405006), and the Center for Computational Science and Engineering at Southern University of Science and Technology.
~\\  

The repository \url{https://github.com/ranwei-ethz/dfs-appendix} contains MATLAB implementations of the Max-Robust-Sum (MRS) and Sum-Robust-Sum (SRS) test statistics, along with illustrative figures demonstrating the alignment of derived probability distribution functions (pdfs) with numerical simulations.  

\clearpage
\nocite{*}
\bibliographystyle{tfs}
\bibliography{dktest}

\clearpage

\section*{Appendix: Supplemental case studies}

\subsection*{Stock returns}

  An issue of debate is if the 1987 stock market crash (Black Monday) was an outlier. We focus on \citep{Schluter}, which is the most recent study on this problem. In \citep{Schluter}, considering daily returns on the Dow Jones Industrial Index, from 3 January 1977 to 31 January 2005, it was claimed that Black Monday is not an outlier. In further detail, the returns were whitened by taking the residuals of a standard AR(1)-GARCH(1,1) model estimated on the returns. Next, the two largest whitened returns $X_{(2)}$ and $X_{(1)}$ were tested as outlying. The test used relies on the GPD approximation (\ref{eq:GPD}) of the tail of the sample, and requires an estimate of the tail parameter $\alpha$. A sample size of $n=732$ was used to estimate $\alpha$. The test statistic $T_r=X_{(r)}/X_{(r+1)}$, comparing $X_{(r)}$ to the previous (next largest) order statistic $X_{(r+1)}$, was used to test if $X_{(2)}$ and $X_{(1)}$ were outlying. Testing outward, with a level of 0.05, neither of these points were identified as outliers.
    
  \begin{figure}[h!]
  \begin{center} 
  \centerline{\includegraphics[width=6cm]{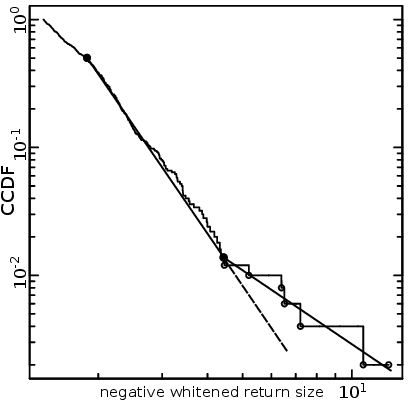}}
  \caption{ \textbf{Stock Returns:} The rough line provides the empirical CCDF of the magnitude of the 500 largest whitened returns of the Dow Jones Industrial Index from 3 January 1977 to 31 January 2005. The solid lines between solid dots provide Pareto model estimates for two magnitude layers. The dashed line extends the slope of the first layer for comparison with that of the second. }
   \label{fig:DKreturns}
  \end{center}
  \end{figure}
    
  To evaluate the approach taken in \citep{Schluter}, we first plot the CCDF of the 500 largest negative whitened returns in Fig.~\ref{fig:DKreturns}. This plot was not provided in \citep{Schluter}, but is clearly essential to assessing above which threshold the GPD approximation (\ref{eq:GPD}) is sound. A few important points are apparent from the figure: Firstly, the CCDF above the $200$ largest observations is shallow/concave, and thus considering more than 200 points (i.e., 732 in \citep{Schluter}) in the sample will weaken the test (i.e., the estimate of $\alpha$ will be too small). Secondly, the second largest point is similar in magnitude to the largest. Thus, the test using $T_1=x_{(1)}/x_{(2)}$ will be masked by $x_{(2)}$, and not rejected. Finally, the top 6 or 7 points seem to follow a heavier tailed distribution. Thus, 6 or 7 points should be tested as outlying, rather than only 2, and a sum test statistic, measuring the cumulative departure of the empirical tail, could be more powerful.
  
  First, we consider estimating a Pareto distribution with two layers. The first layer, containing 193 points, covers $1.97 < x \leq 4.45$ and has MLE $\hat\alpha_1=3.8$. The second layer, containing the 7 largest  points, covers $4.45 < X$ and has MLE $\hat\alpha_2=1.8$. Given that the first layer model is true, there is a $p=0.02$ probability of observing such an extreme difference between the estimated parameters. This two layer model appears to describe the empirical CCDF well (Fig.~\ref{fig:DKreturns}). Next, a single layer model for the top 200 points, covering $4.45 < X$ was estimated with MLE $\hat\alpha_0=3.9$. The likelihood ratio test of the two layer versus one layer model is rejected in favour of the two layer with p-value $0.07$. Further, applying the SS test for $r=6$ with the top $200$ points rejects that there are no outliers with $p=0.04$. Finally, applying the DK test for 6 outliers, for upper sample sizes ranging from 20 to 200, all tests have $p<0.04$. Thus it appears that the 6 largest points are outlying.

  The largest one is, unsurprisingly, `Black Monday' Oct. 19, 1987, which is unambiguously classified as an outlier. An enormous literature has dwelled on its possible origin with a lot of confusion as no simple proximate cause can explain its occurrence. We find more compelling the story that it marked the end of a large financial bubble and thus corresponded to its burst \citep{SJB96,SornetteWhyMarketsCrash,JohansenDrawdown}.
  The second largest event occurred on `Black Friday' Oct. 13, 1989 and is usually associated with 
  a fall of the junk bond market (\url{https://en.wikipedia.org/wiki/Friday_the_13th_mini-crash}).
  The third largest loss corresponds to the first day of reopening of the US stock markets
  on Sept. 17, 2001 after Sept. 11, 2001. It is not clear to us how to interpret the fourth largest loss
  that happened on Nov. 15, 1991. The fifth largest loss on Oct. 27, 1997 is analyzed in details in 
  \citep{SornetteWhyMarketsCrash}, which paints a picture much richer than the usual story
  that this was a global stock market crash caused by an economic crisis in Asia.
  This loss can actually be seen also as a partial burst of a bubble that had been surging in the few
  previous years (recall the famous quip on the `irrational exuberance' of the stock markets by
  Alan Greenspan, then the Chairman of the US Federal Reserve, on Dec. 5, 1996
  (\url{http://www.federalreserve.gov/boarddocs/speeches/1996/19961205.htm})).
  The sixth largest loss on Nov. 9, 1986 is not clearly associated with any exogenous cause,
  to the best of our knowledge. These six outliers are part of the list found by other researchers
  (e.g. \citep{Forture93}).

\subsection*{Nuclear accidents}
  
  We consider as events accidents occurring at nuclear power plants, studied in \citep{WheatleyNuclear2015, Ayoub2021, Ayoub2023}. For this two measures of severity are considered: the cost measured in 2011 US Dollars, for which there are 173 values over the period of 1960 to 2015; and a logarithmic measure of radiation released called the Nuclear Accident Magnitude Scale (NAMS) \citep{Smythe}, for which there are 33 values over the same period. Since the disaster at Fukushima in 2011, Nuclear power has come under major public scrutiny. Further, the level of risk that the nuclear industry claims is consistently much lower than statistical analysis of past events indicates \citep{SornettePSA}. Thus, it is crucial to arrive at a better understanding of the true risk level in this critical application.
  
  The disasters occurring at Chernobyl (1986) and Fukushima (2011) are the most costly accidents thus far, and together are estimated to have caused damage costing 430 Billion 2011 US dollars. This is roughly equal to five times the cost of all 173 other events together.  These events, together with TMI (Three Mile Island, 1979), are also the largest radiation release events. These events are thus extremely large. It is instructive to ask whether a heavy Pareto tail is sufficient to account for these extreme risks or, alternatively, if the tests discussed here can identify outliers / DKs in this data. 

\begin{figure}[h!] 
  \begin{center}
  \centerline{\includegraphics[width=6cm]{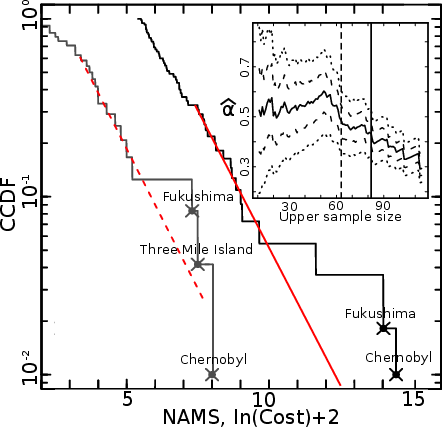}}
  \caption{ \textbf{Nuclear Power Plant Accidents:} the CCDF (cumulative complementary distribution function, i.e., $Pr\lbrace X>x \rbrace$) of the log of the 56 largest cost events, in millions of US Dollars, (black solid) shifted by 2 units for visibility, and the CCDF of all 33 log radiation release (NAMS) values, in solid grey. The fitted lines are exponential MLE fits.
  The inner panel is the Hill plot for the cost values. The solid rough line is the MLE for the exponential distribution for the tail of cost events, for multiple upper sample sizes. It is bracketed by lines indicating one and two standard deviations of the estimator. The vertical dashed line indicates the largest sample at which the exponential cannot be rejected (based on AIC) as being as good as the logspline nonparametric fit. The vertical solid line indicates the largest sample at which the exponential cannot be rejected by the KS test.   }
  \label{fig:nuclear}
  \end{center}
  \end{figure}

  In Fig.~\ref{fig:nuclear} the empirical CCDF (complementary CDF i.e., $Pr\lbrace X>x \rbrace$) for NAMS and the log cost are plotted. For log cost only the 114 events occurring post TMI are included due to an abrupt change in distribution after TMI. For NAMS, the largest three events form a cluster, and appear outlying relative to the exponential distribution with $\hat{\alpha}_{NAMS}=0.7~(0.3)$ fit by MLE to the top 15 points. Not surprisingly the distribution of NAMS and the log cost are similar, as they are certainly related. For log cost, the two or three largest events appear to be outlying relative to the exponential distribution with $\hat{\alpha}_{\$}=0.6~(0.14)$ fit by MLE to the top 50 points. As shown in the Hill plot, inset in Fig.~\ref{fig:nuclear}, when comparing the AIC of the logspline nonparametric fit with the exponential one, the exponential cannot be rejected for samples smaller than the 60 largest points. Further, when performing the KS test, the exponential fit cannot be rejected (at a level of 0.05) for samples smaller than the 80 largest points. Thus the exponential approximation for the tail, and thus the outlier test, should only be applied to not more than the upper 60 points.  
    
  We now test the outliers with a number of the aforementioned tests. The results are presented in Tab.~\ref{tab:nuclear} and summarized below. First considering NAMS, in Fig.~\ref{fig:nuclear} the CCDF is visibly concave until the top 15 points or so, causing a decrease of test power for tests applied to larger sample fractions  (Sec.~2.7). Since the outliers are clustered, (from Sections 2.4 and 2.6) the mixture approach is most powerful, and inward tests the weakest. Despite the small sample size, the mixture test consistently identifies 2 or 3 outliers over a range of upper samples. This confirms that the cluster of large events is a significant feature, however this cluster of large events is not far enough beyond the tail for the other tests to reject the null. It is also important to note that the sample size is very small, and thus our ability to diagnose the validity of the null is weak! Next, cost values are considered for which a larger sample is available. The outward test and the mixture test consistently identify the two largest points as significant outliers. The SRS block test fluctuates around a value of about 0.1. It is not surprising that tests based on the MRS fail to reject due to lack of power when the largest point is not extremely large. Thus Fukushima and Chernobyl appear to be outliers in both radiation released (NAMS) and cost. This is compatible with our understanding of these accidents, where the disaster escalated beyond the threshold of control, leading to an unmitigated proliferation of damage. That these points are outlying in both (dependent) samples would give higher significance if a bi-variate outlier test were performed.
  
  It is worth mentioning that there is a positive relationship between NAMS and cost: Considering the 30 events with substantial radiation release (NAMS$>0$), a linear regression of the logarithm of cost (the response) versus NAMS (the explanatory variable) yields an intercept of $2.33~(0.7),~p=0.003$ and a slope of $0.97~(0.17),~p<10^{-5}$, with coefficient of determination $R^2=0.5$. Further, the same regression can be done for the 16 events that have occurred at Sellafield, in the UK. The result of this is an intercept of $2.30~(1.0),~p=0.04$ and a slope of $1.17~(0.39),~p=0.001$, with coefficient of determination $R^2=0.4$. Thus, there is a significant relationship between radiation release and cost, where we have simply considered a linear relationship. Of course the regression parameters for different plants will depend on the value of property development around the plant.
       
        \begin{table}[!h] 
	\begin{center}
	\scalebox{0.9}{
	\begin{tabular}{l | c  c | c c c c c c c}
	\toprule
	Data&$n$ 	& $r=m$ & MRS   	& SRS    & 	MS Out				&MRS In	  & Mix			& DK 		\\
	\midrule
 	NAMS&$20$ 	& $3$& $0.62$ 		& $0.35$ &		$0,~0.09>0.04$		&$0,~0.62$& $\mathbf{2, 0.03}$		& $0.21$ \\
	NAMS&$15$ 	& $3$& $0.60$ 		& $0.32$&		$0,~0.07>0.04$ 		&$0,~0.60$& $\mathbf{2, 0.025}$		& $0.20 $	 \\
	NAMS&$10$ 	& $3$& $0.37$ 		& $0.15$&		$\mathbf{3,~0.025<0.04}$ &$0,~0.37$& $\mathbf{3, 0.025}$	& $0.14 $      \\
	\midrule
	Damage&$50$ 	& $2$& $0.17$		& $\mathbf{0.08}$ &	$\mathbf{2, 0.03<0.06}$ &$0, 0.17$		& $\mathbf{2, 0.05}$	& $0.18$	 \\
	Damage&$40$ 	& $2$& $0.23$		& $0.11$ 	&	$\mathbf{2, 0.04<0.06}$	&$0, 0.23$		& $\mathbf{2, 0.06}$	& $0.22$	 \\
	Damage&$20$ 	& $2$& $0.25$ 		& $0.14$	 &	$\mathbf{2,~0.05<0.055}$&$0,~0.25$		& $\mathbf{2, 0.07}$	& $0.25$	 \\
	Damage&$15$ 	& $2$& $0.17$ 		& $\mathbf{0.07}$ &	$\mathbf{2,~0.02<0.04}$	&$0,~0.17$		& $\mathbf{2, 0.03}$	& $0.21$	 \\
	Damage&$10$ 	& $2$& $\mathbf{0.06}$ 	& $\mathbf{0.02}$ &	$\mathbf{2,~0.01<0.04}$	&$\mathbf{2,~0.01}$	& $0, 0.18$		& $0.16$	 \\
         \bottomrule
	\end{tabular}	
	}
	\end{center}
	  \caption{Summary of outlier tests for NAMS and cost data for the upper $n$ points, for $r$ outliers (with robustness value $m=r$). Bold values indicate significance at a level of $a=0.1$. Block tests performed include: MRS (\ref{eq:MRS}), SRS (\ref{eq:SRS}), mixture likelihood ratio (\ref{eq:Mix}), and the DK test (\ref{eq:DK}). Further, the MS test was applied outward (MS Out), with the number of identified outliers, the p-value, and the adjusted level (to achieve $a=0.1$) given. For instance, in the first row for MS Out there are zero outliers because the p-value of $0.09$ is above the adjusted level of $0.04$. Finally, the MRS test was applied inward (MRS In), with the number of identified outliers, and the p-value of the test for the largest point given.}
	\label{tab:nuclear}
\end{table}

\subsection*{Fatalities in Epidemics}

We now study the number of fatalities caused by outbreaks of bacterial, viral, and parasitic diseases (epidemics). A dataset for this, with 1,285 events covering the period from 1900 to 2024, was provided by The International Disaster Database (EM-DAT). The dataset excludes, and in some case provides only national fatalities for, pandemic events (spanning multiple countries). Thus the dataset was complemented with Spanish (1918), Asian (1957), and Hong Kong (1968) influenza pandemics, which each caused in excess of 1 million fatalities \citep{Potter}; the 2009 H1N1 `Swine' influenza pandemic, which was estimated to cause upwards of 150,000 fatalities \citep{Simonsen}; and the COVID-19 pandemic, which has led to more than 7 million fatalities globally according to the World Health Orgnization (WHO), as of the writing of this paper. All epidemics and pandemics will be simply referred to as events.

  \begin{figure}[h!]
  \begin{center} 
  \centerline{\includegraphics[width=12cm]{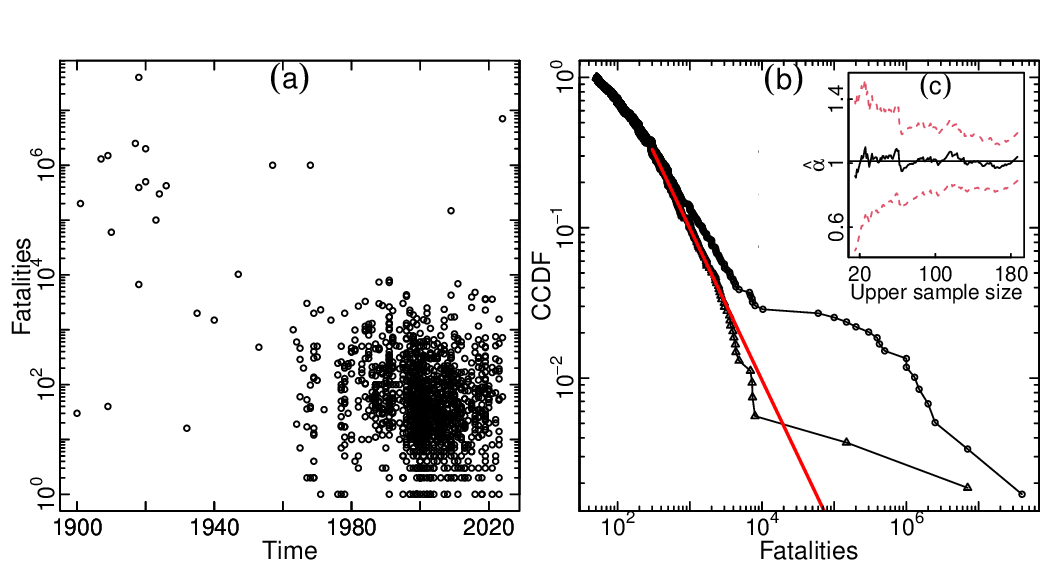}}
  \caption{ \emph{Epidemic Fatalities:} \textbf{(a)} scatterplot of epidemic fatalities from 1900 to 2024. \textbf{(b)} the CCDF of the 593 events in excess of 50 fatalities from 1900-2024 (circles) and the CCDF of the 537 events in excess of 50 fatalities from 1980-2024 (triangles). The latter is fitted with a Pareto tail using a lower threshold $u=300$, and its MLE $\hat\alpha=1.01~(0.03)$ is given by the Hill plot in in panel \textbf{(c)}, where the estimate and associated standard deviation are indicated by the horizontal line and the dash lines. }
   \label{fig:scatter}
  \end{center}
  \end{figure}

From Panel (a) of Fig.~\ref{fig:scatter}, it is clear that over time the dataset has become more complete, in particular for small event sizes. Further, in the period from 1900-1980, 15 events had more than 10,000 fatalities (0.19 per year), whereas in the period from 1980-2024, only 2 such event (H1N1 and COVID) occurred (less than 0.05 per year). Notwithstanding potential changes in the true frequency of events, this is obviously a highly significant difference. These historical extreme events -- Influenzas, Bubonic plagues, Cholera, etc. -- have largely been eradicated through sanitation, vaccines and antibiotics.

 Considering the period from 1900 onwards, many changes have occurred that should have influenced both the incidence and severity of events. Due to data incompleteness, the rate of events cannot be studied. 
 
 Despite this, the sample in excess of 50 fatalities from 1980 onwards, containing 537 points, is roughly stationary in severity. For instance, when repeatedly (1000 times) sampling 100 points from the 537 points, splitting the 100 points into two equal subsamples, and testing their distributions for equivalence with the KS test, only 10.9\% of p-values were less than 0.1. Thus, the modern sample -- spanning the 44 years -- may be used as a proxy to evaluate the outlyingness of the historical extremes, or at least to evaluate how outlying they would be if they were to occur now.
   
 The events in excess of 50 fatalities from both 1900 onwards and 1980 onwards are plotted according to their CCDF in Panel (b) of Fig.~\ref{fig:scatter}. The sample from 1980 approximately has a Pareto tail (see Panel (c)) with parameter around 1.01 (0.03) for the 178 points above the lower threshold of 300. With increasing lower truncations, the estimated parameter increases (as the CCDF bends down), however this is not a significant departure from the estimated tail. For instance, the Anderson-Darling test for the fit of the top 178 points gives a p-value of 0.75. The tail of the sample from 1900 onwards is skewed both by the inclusion of historic large events, and also by the absence of their smaller siblings, which were not recorded. 

The value of the exponent $\alpha \approx 1$ is reminiscent of Zipf's law, which is known to derive quite robustly from the interplay between three simple ingredients \citep{Zipfbook}: birth, proportional growth (also known as `preferential attachment' in network theory) and death. If the variance of the proportional growth component is large, the distribution of event sizes converges to a power law with exponent $\alpha \approx 1$. These ingredients are arguably minimum constituents of epidemic processes and rationalize our finding $\alpha = 1.01~(0.03)$. What is really surprising is the detection of outliers that we present below, which, in some cases, suggests the activation of strong amplification processes beyond the proportional growth mechanisms.
 
We turn our attention to the detection of outliers relative to the approximately stationary data from 1980 onwards. The 17 events in excess of 10,000 fatalities -- 15 of which happened before 1980 -- are considered. The two events closest to the threshold are the Cholera outbreak with 10,276 fatalities (Egypt, 1947) and the pneumonic plague with 60,000 fatalities (China, 1910). Given the smallest event is very close to the threshold, it is treated as a non-outlier for the purposes of this analysis. We begin with the second smallest event, considering as a sample the 176 points with between 300 and 10,000 fatalities occurring since 1980, plus the aforementioned Cholera and pneumonic outbreak. Testing for a single outlier with the DK test (\ref{eq:DK}) gives a p-value of 0. Thus any of the other suspected outliers would be identified as significant outliers too. And, including multiple of these outliers in the sample, and testing them together, would provide equally high significance. 

With respect to the mechanisms at the origin of these outliers, it is likely that each case may be associated with specific catalysing processes. For one of the largest Dragon Kings, the Spanish flu of 1918, serves as a clear example with an identified amplification mechanism. This epidemic infected 500--600 million people, a third of the world's population at that time, and claimed an estimated 40-50 million lives, about five times the toll of the First World War. The first cases of the unknown disease were registered in Kansas, America, in January 1918. By March 1918, more than 100 soldiers fell ill at the US army camp in Funston, Haskell County, where more than 5000 recruits were training for further military operations on the European battlefronts of the First World War. Most of the recruits were farmers, had regular contact with domestic animals and were less resistant to viruses than recruits from cities. The high concentration of personnel in the camp simplified human-to-human transmission. At that time, viruses were not known to medicine, and some doctors had not even accepted the idea that microorganisms could cause disease. Later, the personnel of Funston camp were transferred to Europe by ship, and during the long transatlantic crossing, the virus spread among soldiers coming from other parts of the USA. Upon arriving in Europe, American soldiers infected British and French forces, which in their turn infected German forces in hand-to-hand combat. When Woodrow Wilson, President of the United States from 1913 to 1921, began to receive reports about a severe epidemic among American forces, he made no public acknowledgement of the disease \citep{Barry2004}. Moreover, other governments involved in the war made similar decisions -- censorship, lies, and even active propaganda -- to keep up morale, allowing the disease to continue to spread without any preventive measures. The pandemic was named `Spanish flu' because Spain was a neutral country during the First World War and did not suppress the media, so it was only Spanish newspapers that published honest articles about the severity of the disease -- despite the fact that it had originated in the USA and spread initially among American soldiers in the absence of a proper response by the US government. This lack of response was probably due to the US strategic goal of developing a strong political influence in the post-WWI peace process that was to shape international politics in the following decades. In summary, the amplification of the pandemic can be attributed to two key mechanisms: the highly efficient transmission facilitated by the movement of soldiers and the absence of any preventive measures due to the war's priority.

Similarly, the COVID-19 pandemic reflects the characteristics of an amplification mechanism, where specific catalyzing factors led to an unprecedented global impact. The virus, SARS-CoV-2, first emerged in Wuhan, China, in late 2019 and rapidly spread worldwide. The pandemic was exacerbated by factors such as global interconnectedness, delays in implementing public health measures, and varying levels of preparedness among countries. By mid-2024, the pandemic had resulted in over 7 million confirmed deaths, with actual figures likely higher due to underreporting and discrepancies in data collection. Economies around the world were severely impacted as measures such as lockdowns and social distancing, designed to contain the virus, led to unprecedented disruptions in global supply chains and labor markets. Unlike the Spanish flu, where war efforts took precedence, the initial underestimation of COVID-19 severity by governments and the slow international response played significant roles in its spread. The COVID-19 pandemic highlights the potential for future events in the realm of global health, where specific conditions -- such as increased travel, urbanization, and environmental degradation -- can lead to catastrophic outcomes. This underscores the importance of recognizing and mitigating the factors that contribute to such extreme events, particularly in the context of ongoing threats like antimicrobial resistance and climate change. 
 
We thus conclude that we found evidence of Dragon Kings in the database of epidemic events, with the 2009 swine flu pandemic and the more extreme COVID-19 being notable examples in the more recent post-1980 period. Although the AIDS epidemic was not included in this analysis, it represents another significant outlier in the realm of public health. In 2014, 1.2 million [1 million--1.5 million] people died from AIDS-related illnesses, a marked improvement from the peak in 2015, when 2.3 million [2.1 million--2.6 million] deaths were recorded. Since its identification, an estimated $\sim 36$ million people have died from AIDS-related causes \citep{UNAIDS1,UNAIDS2}. The concept of Dragon King in epidemic dynamics suggests that while such extreme events are rare, they are not beyond the realm of possibility, especially in a world where human activities increasingly intersect with natural processes in unpredictable ways.

 \subsection*{City sizes \label{cityDKsect}}
  
    \begin{figure}[h!] 
  \begin{center}
  \centerline{\includegraphics[width=6cm]{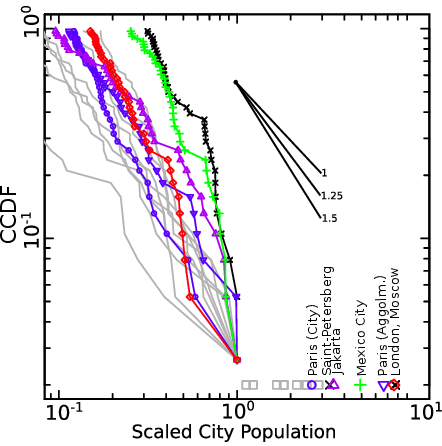}}
  \caption{ \textbf{City sizes}: plot of the CCDF for the 35 largest cities (and also agglomerations for France) in each of the 14 countries: Brasil, China, France, India, Indonesia, Japan, Korea, Mexico, Nigeria, Pakistan, Phillipines, Russia, the UK, and the USA. The sizes were scaled such that the second largest point (third largest for Russia) in each country has size 1. The scaled largest point (two largest for Russia) are plotted in the bottom right. Each country that is suspected of having outliers is in colour: France (blue circles for cities, blue downward triangles for agglomerations), Russia (black x marks), Indonesia (purple triangles), Mexico (green crosses), and England (red squares). }
  \label{fig:cities}
  \end{center}
  \end{figure} 
      
  Within the disciplines of economics, geography and geopolitics (among others), 
  the distribution of city and of agglomeration sizes is of particular interest, due to the importance of urban primacy, and because it constitutes one of the key stylized facts. There is a large literature documenting that the distributions of city and agglomeration sizes follows a Pareto distribution with parameter close to one (Zipf's Law) (see e.g. \citep{Zipfbook} and references therein). There has been some debate over if the distribution would be better represented by a log-normal \citep{Eeckhout2004,Eeckhout2008,Levy2007}, however the debate has been clearly settled in favour of the Pareto for the 1000 largest cities \citep{Malev2011}. Note that both the Pareto and log-normal are generally taken as result from Gibrat's principle of proportional growth \citep{Gibrat1931}
  (see \citep{Zipfbook} for a general derivation).

  In \citep{DKpisarenko11}, the DK test (\ref{eq:DK}) was used to identify outlying population agglomerations for a number of countries, assuming a Pareto tail. Here we consider city sizes rather than agglomerations since this data is available for more countries. We only consider agglomeration sizes for the case of Paris, France for comparison with \citep{DKpisarenko11}. Data for 14 large countries\footnote{Brasil, China, France, India, Indonesia, Japan, Korea, Mexico, Nigeria, Pakistan, Phillipines, Russia, the UK, and the USA.} were taken from \cite{citysizes}. All tests use the SRS block test statistic for testing the largest point as an outlier, with the exception or Russia where two outliers are tested.
      
  In Fig.~\ref{fig:cities}, the 35 largest cities of each country are plotted according to their empirical CCDF, rescaled in a way to make the largest cities comparable. Since not all of the samples appear to follow a pure Pareto, results on robustness and testing the tail (Sections \ref{sec:robust} and \ref{sec:general}) are relevant here.
%
%
  First considering French cities, for upper sample sizes of $5<n\leq35$, the p-value fluctuates in a range of $0.1-0.2$. Thus, there is only marginal evidence that the city of Paris is an outlier. However, the agglomeration of Paris is relatively larger, and for $5<n\leq25$ the p-value fluctuates between 0.02 and 0.15, providing stronger evidence of the uniqueness of Paris. The CCDF of Indonesia is concave. Thus, if too large of a sample is considered in the test, Jakarta will not be detected as an outlier. For instance, if one draws a line that best interpolates all points of the empirical CCDF, the line will be so shallow that the Jakarta point falls beneath it, essentially masking the outlier. For this reason, Jakarta, Indonesia has $p<0.1$ only for the upper most points $5<n<11$. Mexico is an even more extreme case of the above, having $p<0.1$ for $5<n<20$ for Mexico City. London, UK, is the most significant, having $0.001<p<0.05$ for all $5<n\leq35$. Finally, testing both Moscow and Saint-Petersberg as outliers, the p-value is in  $0.01<p<0.15$, with a mean of $0.09$ for all $5<n\leq35$. In conclusion, it is absolutely clear that London is an outlier, and the largest city/cities of five of the remaining fourteen countries considered have moderate/suggestive evidence that they are outlying. A plausible mechanism for the outlier status of London (and other cities) can be attributed to the positive feedback loops that have bolstered the outsized importance of these imperial power centers over the past centuries and decades, coupled with their self-reinforcing economic attractiveness.

\end{document}